\documentclass[journal]{IEEEtran}
\usepackage{slashbox}
\usepackage{graphicx}
\usepackage{citesort}
\usepackage{amssymb}
\usepackage{amsfonts}
\usepackage{amsmath}
\usepackage{epsfig}
\usepackage{fancybox}
\usepackage{textcomp}
\usepackage{multirow}
\usepackage{booktabs}
\usepackage{mathrsfs}
\usepackage{pifont}
\usepackage[ruled,vlined,linesnumbered]{algorithm2e}
\usepackage{euscript}

\usepackage[dvipsnames,usenames]{color}
\usepackage[usenames,dvipsnames]{color}

\newcommand{\argmax}{\operatornamewithlimits{arg\, max}}

\usepackage{tabularx}
\newcolumntype{L}[1]{>{\raggedright\arraybackslash}p{#1}}
\newcolumntype{C}[1]{>{\centering\arraybackslash}p{#1}}
\newcolumntype{R}[1]{>{\raggedleft\arraybackslash}p{#1}}

\begin{document}

\title{State Estimation in Smart Distribution System With Low-Precision Measurements}

\author{Jung-Chieh~Chen,~\IEEEmembership{Member,~IEEE}, Hwei-Ming~Chung, Chao-Kai~Wen,~\IEEEmembership{Member,~IEEE}, Wen-Tai~Li,~\IEEEmembership{Student Member,~IEEE}, and~Jen-Hao~Teng,~\IEEEmembership{Senior Member,~IEEE}
\thanks{J.-C.~Chen is with the Department of Optoelectronics and Communication Engineering, National Kaohsiung Normal University, Kaohsiung 80201, Taiwan (e-mail: $\rm
jcchen@nknu.edu.tw$).}
\thanks{H.-M.~Chung is with the Research Center for Information Technology Innovation, Academia Sinica, Taipei 11529, Taiwan.}
\thanks{C.-K.~Wen is with the Institute of Communications Engineering, National Sun Yat-sen University, Kaohsiung 80424, Taiwan.}
\thanks{W.-T.~Li is with Engineering Product Development, Singapore University of Technology and Design, Singapore.}
\thanks{J.-H. Teng is with the Department of Electrical Engineering, National Sun Yat-sen University, Kaohsiung 80424, Taiwan.}

}

\markboth{IEEE Access}%
{Shell \MakeLowercase{\textit{et al.}}: Bare Demo of IEEEtran.cls
for Journals}

\maketitle

\begin{abstract}
Efficient and accurate state estimation is essential for the optimal management of the future smart grid.
However, to meet the requirements of deploying the future grid at a large scale, the state estimation algorithm must be able to accomplish two major tasks:
(1) combining measurement data with different qualities to attain an optimal state estimate and
(2) dealing with the large number of measurement data rendered by meter devices.
To address these two tasks, we first propose a \emph{practical} solution using a very short word length to represent a partial measurement of the system state in the meter device to reduce the amount of data.
We then develop a unified probabilistic framework based on a Bayesian belief inference to incorporate measurements of different qualities to obtain an optimal state estimate.
Simulation results demonstrate that the proposed scheme \emph{significantly} outperforms other linear estimators in different test scenarios.
These findings indicate that the proposed scheme not only has the ability to integrate data with different qualities but can also decrease the amount of data that needs to be transmitted and processed.

\end{abstract}


\section{Introduction}

Integrating renewable energy generations, distributed
microgenerators, and storage systems into power grids is one of the
key features of enabling the future smart grid \cite{SG-techreport}.
However, this integration gives rise to new challenges, such as the
appearance of overvoltages at the distribution level. Accurate and
reliable \emph{state estimation} must be developed to achieve the
real-time monitoring and control of this hybrid distributed
generation system and therefore assure the proper and reliable
operation of the future grid \cite{Huang-12}. An increase in the
penetration of the distributed generator necessarily leads to an
unusual increase in measurements \cite{Nagasawa-12}.

Conventional state estimation techniques, such as the weighted least
squares (WLS) algorithm, rely on measurements from the supervisory
control and data acquisition (SCADA) systems \cite{abur2004powe}. A
well-known fact is that the measurements provided by SCADA are
intrinsically \emph{less} accurate \cite{Li-14-TPS,Gol-14}.
Moreover, adapting conventional WLS technique to SCADA-based state
estimation is \emph{not} robust due to its vulnerability to poor
measurements \cite{Gol-14}. More recently, the deployment of
high-precision phasor measurement units (PMUs) in electric power
grids has been proven to improve the accuracy of state estimation
algorithms \cite{Hurtgen-08,Phadke-09,PMU-09-book,Gol-14}. However,
PMUs remain expensive at present, and \emph{limited} PMU
measurements, along with conventional SCADA measurements, must be
incorporated into the state estimator for the active control of the
smart grid.

Several state estimation methods using a mix of conventional SCADA and PMU measurements have already been proposed for electric power grids, as shown in Refs. \cite{Zhou-06,Gol-15}.
However, the joint processing of measurements of \emph{different} qualities may result in an \emph{ill-conditioned} system.
Moreover, another critical challenge but essential task in deploying the future grid at a large scale is the \emph{massive} amount of measurement data that needs to be transmitted to the data processing control center (DPCC).
This poses a risk to the grid's operator: DPCC is drowning in data overload, a phenomenon called ``data tsunami.''
A massive amount of measurement data also results in a long time for data collection, so that the state estimation result is not prompt.
To alleviate the impact of data tsunami, Alam et al. \cite{Alam-14} took advantage of the compressibility of spatial power measurements to decrease the number of measurements based on the compressive sensing technique.
Nevertheless, the performance of \cite{Alam-14} is relatively sensitive to the influence of the so-called compressive sensing matrix.

We first propose a practical solution to address the abovementioned
challenges. Inspired by \cite{Alam-14}, we can compress the
measurement not only with compressive sensing matrix but also
itself. Therefore, we use a very short length to compress the
partial measurements of the system.\footnote{The work in
\cite{Alam-14} designed a compressed matrix to shorten the
measurements, where the compressed measurements are \emph{still}
represented with 12 or 16 bits for transmission. However, in the
present study, partial measurements are represented in extremely
short length for transmission. Therefore, the focus of our study is
different from that of \cite{Alam-14}.} The use of a \emph{very
short} word length (e.g., 1-6 bits)\footnote{In practical
application, all of the measurements obtained by the meter devices
must be quantized before being transmitted to the DPCC for
processing. Modern SCADA systems use a typical word length of 12 (or
16) bits to represent the measurements employed to obtain a
high-resolution quantized measurement.} to represent a partial
measurement of the system state in the meter device reduces the
amount of data that the DPCC needs to process. This data-reduction
technique considerably enhances the efficiency of the grid's
communication infrastructure and bandwidth because only a limited
number of bits representing the measurements are sent to the DPCC.
In addition, instead of substituting all sensors in the cerrunt
power system with PMUs, we only have to add several wireless meters
with low bit analog-to-digital converter, which are cheaper than
conventional meters. Hence, the cost of placing the meters can be
reduced.

Nevertheless, the traditional state estimation methods cannot be
applied to the system with partial measurements represented by very
short length. Thus, we develop a new scheme to obtain an optimal
state estimate and then minimize the performance loss due to
quantization while incorporating measurements of different
qualities. Before designing the state estimation algorithm, we first
formalize the linear state estimation problem using data with
different qualities as a probabilistic inference problem. Then, this
problem can be tackled efficiently by describing an appropriate
factor graph related to the power grid topology. Particularly, the
factorization properties of the factor graphs improve the accuracy
of mixing measurements of different qualities. Then, the concept of
the estimation algorithm is motivated by using the maximum
posteriori (MAP) estimate to construct the system states.

The proposed MAP state estimate algorithm derived from the generalized approximate message passing (GAMP)-based algorithms \cite{Rangan-11-ISIT,Krzakala-12,SwAMP-2015}, which exhibit excellent performance in terms of both precision and speed in dealing with high-dimensional inference problems, while preserving low complexity.
In contrast to the traditional \emph{linear} solver for state estimation, which does not use prior information on the system state, the proposed scheme can learn and therefore exploit prior information on the system state by using expectation-maximization (EM) algorithm \cite{Vila-13TSP} based on the estimation result for each iteration.

The proposed framework is tested in  different test systems.
The simulation results show that the proposed algorithm performs \emph{significantly} better than other linear estimates.
In addition, by using the proposed algorithm, the obtained state estimations retain accurate results, even when \emph{more than half} of the measurements are quantized to a very short word length.
Thus, the proposed algorithm can integrate data with different qualities while reducing the amount of data.

{\em Notations}---Throughout the paper, we use $\mathbb{R}$ and $\mathbb{C}$ to represent the set of real numbers and complex numbers, respectively.
The superscripts $(\cdot)^{\mathsf{H}}$ and $(\cdot)^{*}$ denote the Hermitian transposition and conjugate transpose, respectively.
The identity matrix of size $N$ is denoted by $\mathbf{I}_{N}$ or simply $\mathbf{I}$.
A complex Gaussian random variable $x$ with mean $\widehat{x}$ and variance $\sigma^2_{x}$ is denoted by $\mathscr{N}_{\mathbb{C}}(x;\widehat{x},\sigma^2_{x}) \triangleq (\pi\sigma^2_{x})^{-1}\exp(-|x-\widehat{x}|^2/\sigma^2_{x})$ or simply $\mathscr{N}_{\mathbb{C}}(\widehat{x},\sigma^2_{x})$.
$\mathbb{E}[\cdot]$ and $\mathbb{VAR}[\cdot]$ represent the expectation and variance operators, respectively.
$\Re\{ \cdot \}$ and $\Im\{ \cdot \}$ return the real and imaginary parts of its input argument, respectively.
$\arg(\cdot)$ returns the principal argument of its input complex number.
Finally, $\mathrm{j} \triangleq \sqrt{-1}$.

\section{System Model and Data Reduction}\label{sec:02}

\subsection{System Model}

\begin{figure}
\begin{center}
\resizebox{3.35in}{!}{%
\includegraphics*{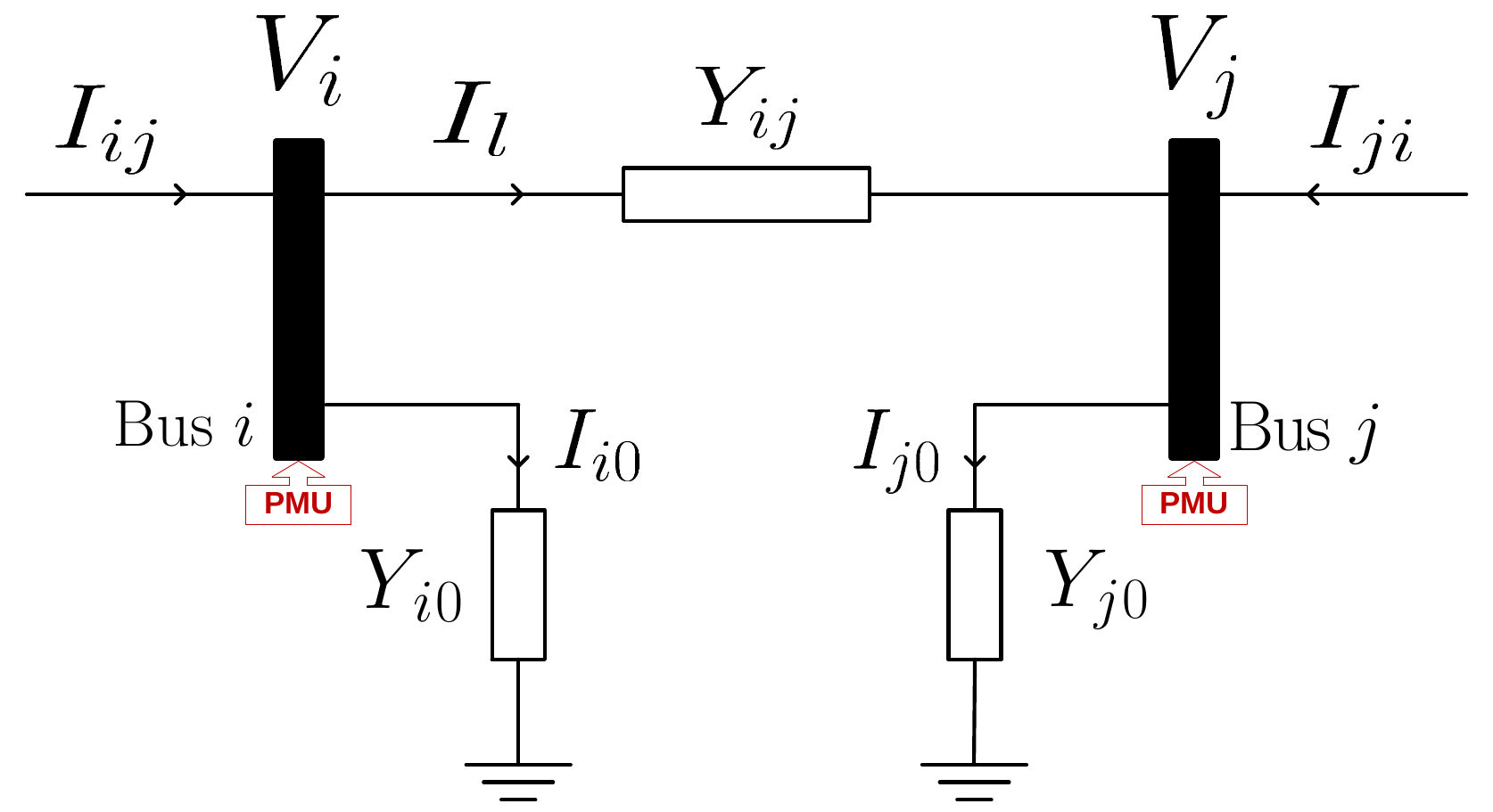} }%
\caption{Transmission $\pi$-line model for calculating line flows.}\label{fig:line_ex}
\end{center}
\end{figure}

Our interest is oriented toward applications in the distribution system.
Following the canonical work on the formulation of the linear state estimation problem \cite{PMU-09-book} and power flow analysis \cite{Chen-91-TPD}, we use a $\pi$-model transmission line to indicate how voltage and current measurements are related to the considered linear state estimation problem.
For easy understanding of this model, we start with a $\pi$-equivalent of a transmission line connecting two PMU-equipped buses $i$ and $j$ as shown in Fig.\ref{fig:line_ex}, where $Y_{ij}$ is the series admittance of the
transmission line, $Y_{i0}$ and $Y_{j0}$ are the shunt admittances of the side of the transmission line in which the current measurements $I_{i0}$ and $I_{j0}$ are taken, respectively, and the parallel conductance is neglected.
In this case, the system state variables are the voltage magnitude and angle at each end of the transmission line, that is, $V_{i}\in \mathbb{C}$ and $V_{j}\in \mathbb{C}$.

In Fig. \ref{fig:line_ex}, the line current $I_{ij}$, measured at bus $i$, is positive in the direction flowing from bus $i$ to bus $j$, which is given by
\begin{equation}\label{eq:I_ij}
    I_{ij} = I_{l}+I_{i0} =Y_{ij}\left(V_{i}-V_{j}\right) +Y_{i0}V_{i}.
\end{equation}
Likewise, the line current $I_{ji}$, measured at bus $j$, is positive in the direction flowing from bus $j$ to bus $i$, which can be expressed as
\begin{equation}\label{eq:I_ji}
    I_{ji} = -I_{l}+I_{j0} =Y_{ij}\left(V_{j}-V_{i}\right) +Y_{j0}V_{j}.
\end{equation}
Then, (\ref{eq:I_ij}) and (\ref{eq:I_ji}) can be written in matrix form as
\begin{equation}\label{eq:transmission_line__matrix}
    \begin{bmatrix}
     I_{ij} \\
     I_{ji}
    \end{bmatrix}
     =    \begin{bmatrix}
      Y_{ij}+Y_{i0}  & -Y_{ij}\\
      -Y_{ij} &  Y_{ij}+Y_{j0}
    \end{bmatrix}
    \begin{bmatrix}
     V_{i} \\
     V_{j}
    \end{bmatrix}.
\end{equation}
Given that PMU devices are installed in both buses, the bus voltage and the current flows through the bus are available through PMU measurements.
Based on these measured data, the \emph{complete} state equation can be expressed as
\begin{equation}\label{eq:basic_linear_eq}
    \begin{bmatrix}
            V_{i}\\
            V_{j}\\
            I_{ij}\\
            I_{ji}
    \end{bmatrix}
     =   \underbrace{ \begin{bmatrix}
      1 & 0 \\
            0 & 1 \\
            Y_{ij}+Y_{i0} & -Y_{ij} \\
            -Y_{ij} & Y_{ij}+Y_{j0}
    \end{bmatrix}}_{\triangleq\, {\bf H}}
    \begin{bmatrix}
     V_{i} \\
     V_{j}
    \end{bmatrix}.
\end{equation}
Here, ${\bf H}$ can be decomposed into four matrices related to power system topology \cite{PMU-09-book,Jones-11}.
These matrices are termed the current measurement-bus incidence matrix, the voltage measurement-bus incidence matrix, the series admittance matrix, and the shunt admittance matrix, as explained later in this section.
Thereafter, (\ref{eq:basic_linear_eq}) can be further extended to the general model in power systems.

\begin{figure}
\begin{center}
\resizebox{3.5in}{!}{%
\includegraphics*{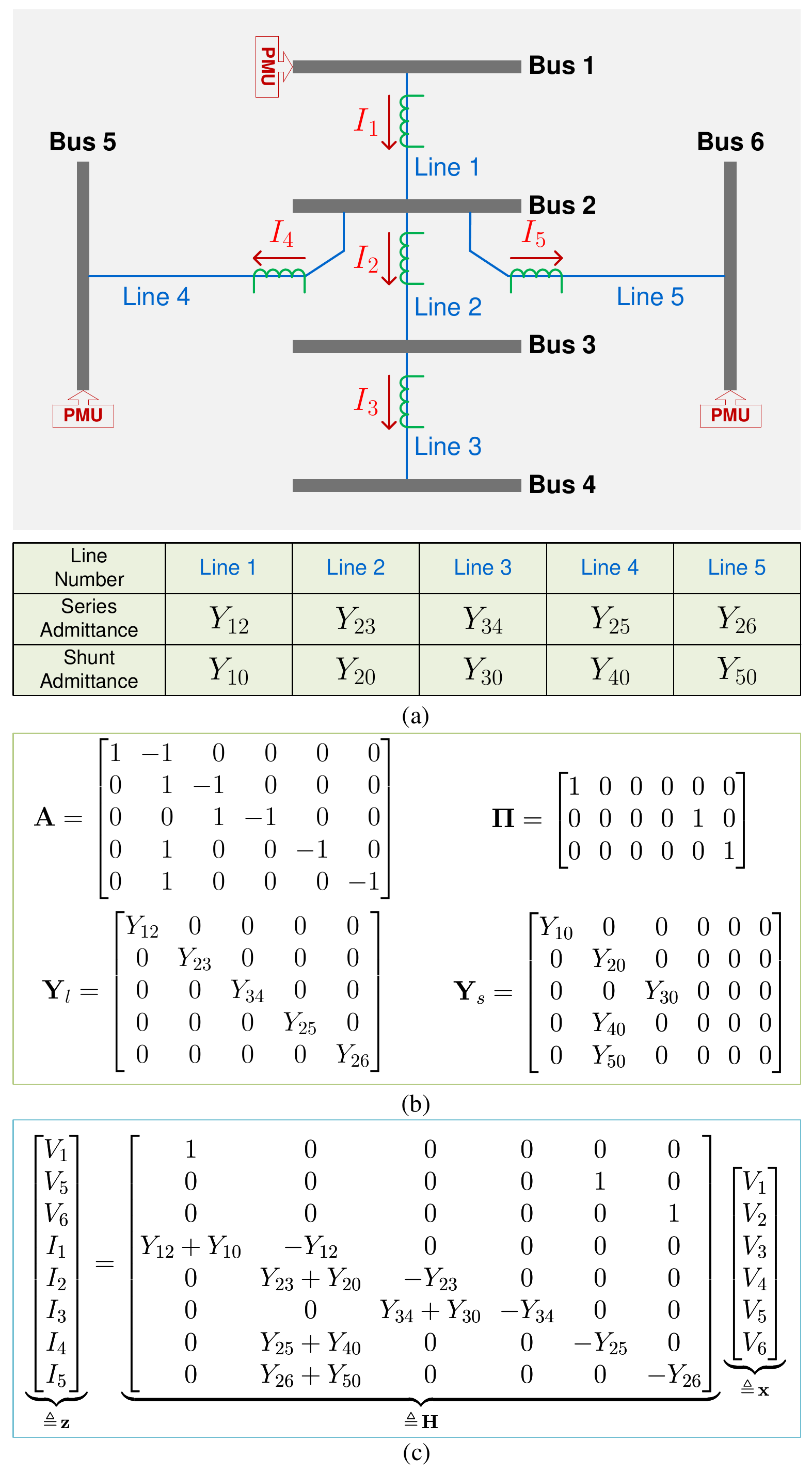} }%
\caption{(a) A fictitious six-bus system, where current measurement is represented by an arrow above a current meter. (b) The corresponding $\mathbf{A}$, $\boldsymbol{\Pi}$, $\mathbf{Y}_{l}$,
and $\mathbf{Y}_{s}$ for this six-bus system. (c) The full state equation for this six-bus system.}\label{fig:system_model_example}
\end{center}
\end{figure}

Before explaining the rules for constructing these matrices used in the state equation, a \emph{fictitious} six-bus system is first presented, as shown in Fig. \ref{fig:system_model_example}(a).
This simple system is used to demonstrate how each of these matrices is constructed for the sake of clarity.
As Fig \ref{fig:system_model_example}(a) indicates, the line current flowing through each line is directly measurable with a current meter.
However, bus voltages are measurable only at buses 1, 5, and 6 because of the PMUs are installed only in these three buses.
Thus, in this example, the number of buses is $N=6$, the number of PMUs (or the number of buses that have a voltage measurement) is $L=3$, and the number of current measurements is $M = 5$.

The explicit rules for constructing each of the four matrices are provided.
First, $\mathbf{A} \in \mathbb{R}^{M\times N}$ is the current measurement-bus incidence matrix that indicates the location of the current flow measurements in the network, where the rows and columns of $\mathbf{A}$ represent, respectively, the \emph{serial} number of the current measurement and the bus number.
More specifically, the entries of $\mathbf{A}$ are defined as follows.
If the $m$-th current measurement $I_{m}$ (corresponding to the $m$-th row) leaves from the $n$-th bus (corresponding to the $n$-th column) and heads toward the $n'$-th bus (corresponding to the $n'$-th column), the $(m,n)$-th element of $\mathbf{A}$ is $1$, the $(m,n')$-th element of $\mathbf{A}$ is $-1$, and all the remaining entries of $\mathbf{A}$ are identically zero.
Second, $\boldsymbol{\Pi} \in \mathbb{R}^{L\times N}$ is the voltage measurement-bus incidence matrix that points out the relationship between a voltage measurement and its corresponding location in the network, where the rows and columns of $\boldsymbol{\Pi}$ represent the \emph{serial} number of the voltage measurement and the bus number, respectively.
Hence, the entries of $\boldsymbol{\Pi}$ can be defined in such a way.
If the $l$-th voltage measurement (corresponding to the $l$-th row) is located at the $n$-th bus (corresponding to the $n$-th column), then the $(l,n)$-th element of $\boldsymbol{\Pi}$ is $1$, and all the other elements of $\boldsymbol{\Pi}$ are zero.
Third, $\mathbf{Y}_{l} \in \mathbb{C}^{M\times M}$, which denotes the series admittance matrix, is a \emph{diagonal} matrix, whose diagonal terms are the line admittance of the transmission line being measured.
Thus, $\mathbf{Y}_{l}$ is populated using the following single rule.
For the $m$-th current measurement, the $(m,m)$-th element of $\mathbf{Y}_{l}$ is the series admittance of the line being measured.
Fourth, $\mathbf{Y}_{s} \in \mathbb{C}^{M\times N}$ is the shunt admittance matrix whose elements are determined by the shunt admittances of the lines which have a current measurement.
The following rules are used to populate the matrix.
If the $m$-th current measurement \emph{leaves} the $n$-th bus, then the $(m,n)$-th element of $\mathbf{Y}_{s}$ is the shunt admittance of the line, and all the other elements of $\mathbf{Y}_{s}$ are zero.
By following these rules, the constructions of $\mathbf{A}$, $\boldsymbol{\Pi}$, $\mathbf{Y}_{l}$, and $\mathbf{Y}_{s}$ for the six-bus system are illustrated in Fig. \ref{fig:system_model_example}(b).

Given the above definitions, the linear state equation in (\ref{eq:basic_linear_eq}) can be further extended to \emph{general} linear state equation with $N$ buses, $L$ voltage measurements, denoted by $\mathbf{v}\in \mathbb{C}^{L}$, and $M$ current measurements, denoted by $\mathbf{i}\in \mathbb{C}^{M}$, as follows \cite{Jones-11}
\begin{equation}\label{eq:basic_linear_matrix_eq_error_free}
    \underbrace{  \begin{bmatrix}
                  \mathbf{v} \\
                  \mathbf{i}
                 \end{bmatrix}}_{\triangleq\, \mathbf{z}}=
    \underbrace{\begin{bmatrix}
              \boldsymbol{\Pi}    \\
              \mathbf{Y}_{l} \mathbf{A} + \mathbf{Y}_{s}
    \end{bmatrix}}_{\triangleq\,  \mathbf{H} }    \mathbf{x},
\end{equation}
where $\mathbf{z}\in \mathbb{C}^{L+M}$ denotes a vertical
concatenation of the set of voltage and current phasor measurements,
$\mathbf{x}\in \mathbb{C}^{N}$ is the \emph{complex} system state,
and ${\bf H}\in \mathbb{C}^{(L+M)\times N}$ is a topology matrix
(i.e., also referred to as the measurement matrix in a general
linear system).\footnote{Using slight modifications, the system
model in (\ref{eq:basic_linear_matrix_eq_error_free}) can easily be
extended to \emph{three-phase} power systems \cite{Jones-11}. Each
element of ${\bf H}$ is modified as follows. Elements ``1'' and
``0'' in $\boldsymbol{\Pi} $ and $\mathbf{A}$ are replaced with a $3
\times 3$ identity matrix and a $3 \times 3$ null matrix,
respectively. Each diagonal element of $\mathbf{Y}_{l}$ is replaced
with $3 \times 3$ admittance structures, whereas the off-diagonal
elements become $3 \times 3$ zero matrices. Finally, each nonzero
element of $\mathbf{Y}_{s}$ is replaced with $3 \times 3$ admittance
structures and the remaining elements become $3 \times 3$ zero
matrices.}

Considering again the fictitious six-bus system presented earlier,
the full system state for this system is also provided in Fig.
\ref{fig:system_model_example}(c) for ease of understanding.
Defining $P\triangleq L+M$ and accounting for the measurement error
in the linear state equation,
(\ref{eq:basic_linear_matrix_eq_error_free}) then
becomes\footnote{As described in (\ref{eq:basic_linear_matrix_eq}),
the considered system model is
 expressed as $\mathbf{y} =  \mathbf{H}  \mathbf{x}  \,+\,
 \mathbf{e}$  because we aimed to estimate $\mathbf{x}$. Thus,
 $\mathbf{H}$ should be at least a square matrix or an
 overdetermined system. In this case, $P = (L+M) \geq N$.}
\begin{equation}\label{eq:basic_linear_matrix_eq}
   \mathbf{y} =  \underbrace{\mathbf{H}  \mathbf{x}}_{=\, \mathbf{z}}  \,+\, \mathbf{e},
\end{equation}
where $\mathbf{y}\in \mathbb{C}^{P}$ is the raw measurement vector
of the voltage and current phasors, $\mathbf{z}\in \mathbb{C}^{P}$
is also referred to as the \emph{noiseless} measurement vector, and
$\mathbf{e}\in \mathbb{C}^{P}$ is the measurement error, in which
each entry is modeled as an identically and independently
distributed (i.i.d.) complex Gaussian random variable with zero mean
and variance $\sigma^{2}$.

\subsection{Data Reduction and Motivation}

In reality, all of the measurements must be quantized before being transmitted to the DPCC for processing.
For example, modern SCADA systems are equipped with an analog device that converts the measurement into binary values (i.e., the usual word lengths are 12 to 16 bits).
To achieve this, the measurements $\mathbf{y}= \left\{y_{\mu}\right\}_{\mu =1}^{P}$ are processed by a complex-valued quantizer in the following componentwise manner:
\begin{equation} \label{eq:Q_y}
              \widetilde{\mathbf{y}} = \left\{\widetilde{y}_{\mu}\right\}_{\mu =1}^{P}= \mathcal{Q}_{\mathbb{C}}\left(\mathbf{y} \right) = \left\{\mathcal{Q}_{\mathbb{C}}\left(y_{\mu} \right) \right\}_{\mu =1}^{P},
\end{equation}
where $\widetilde{\mathbf{y}}=\left\{\widetilde{y}_{\mu}\right\}_{\mu =1}^{P}$ is the \emph{quantized version} of $\mathbf{y}=\left\{y_{\mu}\right\}_{\mu =1}^{P}$, and each complex-valued quantizer $\mathcal{Q}_{\mathbb{C}}\left(\cdot\right)$ is defined as $\widetilde{y}_{\mu} = \mathcal{Q}_{\mathbb{C}}\left(y_{\mu} \right) \triangleq \mathcal{Q}\left(\Re\left(y_{\mu} \right)\right) + \mathrm{j}\mathcal{Q}\left(\Im\left(y_{\mu} \right)\right)$.
This means that, for each complex-valued quantizer, two real-valued quantizers exist that \emph{separately} quantize the real and imaginary part of the measurement data.
Here, the real-valued quantizer $\mathcal{Q}\left(\cdot\right)$ is a $B$ bit midrise quantizer \cite{Proakis-book} that maps a real-valued input to one of $2^B$ \emph{disjoint} quantization regions, which are defined as $\mathscr{R}_{1}=(-\infty,r_{1}], \mathscr{R}_{2}=(r_{1},r_{2}], \ldots, \mathscr{R}_{b}=(r_{b-1},r_{b}], \ldots, \mathscr{R}_{2^{B}}=(r_{2^{B}-1},\infty)$, where $-\infty<r_{1} < r_{2}<\cdots<r_{2^{B}-1}<\infty$.
All the regions, except for $\mathscr{R}_{1}$ and $\mathscr{R}_{2^{B}}$, exhibit equal spacing with increments of $\Delta$.
In this case, the boundary point of $\mathscr{R}_{b}$ is given by $r_{b} = \left( -2^{B-1} + b \right)\Delta$, for $b=1,\ldots, 2^{B}-1$.
Thus, if a real-valued input falls in the region $\mathscr{R}_{b}$, then the quantization output is represented by $r_{b}-\frac{\Delta}{2}$, that is, the \emph{midpoint} of the quantization region in which the input lies.

When the DPCC receives the quantized measurement vector $\widetilde{\mathbf{y}}$, it can perform state estimation using the
linear minimum mean square error (LMMSE) method:
\begin{equation} \label{eq:LMMSE_def}
 \widehat{\mathbf{x}}^{\mathsf{LMMSE}} = \left({\bf H}^{\mathsf{H}} {\bf H} + \sigma^{2} \mathbf{I}\right)^{-1} {\bf H}^{\mathsf{H}}\widetilde{\mathbf{y}}.
\end{equation}
As can be observed, the accuracy of the LMMSE state estimator highly depends on the quantized measurements $\widetilde{\mathbf{y}}$.
A relatively high-resolution quantizer must be employed in the meter device to maintain the high-precision measurement data and therefore prevent the LMMSE performance from being affected by lower-resolution measurements.
However, this is unfortunately accompanied by a significant increase in the data for transmission and processing.
This unusual trend of increasing data motivates the need for a data-reduction solution.

To reduce the amount of high-precision measurement data, we propose
quantizing and representing \emph{partial} measurements using a
\emph{very short} word length (e.g., 1-6 bits), instead of adopting
a higher number of quantization bits to represent \emph{all} the
measurements. In this way, a more efficient use of the available
bandwidth can be achieved. However, lower-resolution measurements
tend to degrade the state estimation performance. Moreover,
quantized measurements with different resolutions require a proper
design of the data fusion process to improve the state estimation
performance. Given the above problems, we develop in the next
section a new framework based on a Bayesian belief inference to
incorporate the quantized measurements from the meter devices
employing different resolution quantizers to obtain an optimal state
estimate.

\section{State Estimation Algorithm}\label{sec:03}
\subsection{Theoretical Foundation and Factor Graph Model}
The objective of this work is to estimate the system state
$\mathbf{x}=\{x_{i}\}_{i=1}^{N}$ from the quantized measurement
vector $\widetilde{\mathbf{y}}$ and the knowledge of matrix ${\bf
H}$ using the minimum mean square error (MMSE) estimator. A
well-known fact is that the Bayesian MMSE inference of $x_i$ is
equal to the posterior mean,\footnote{In what follows, we will
derive the posterior mean and variance based on the MMSE
estimation.} that is,
\begin{equation}\label{eq:compuate_marginal}
      \widehat{x}_{i}^{\mathsf{MMSE}} = \int \,   x_{i}\mathscr{P}(x_{i}|{\bf H} ,\widetilde{\mathbf{y}}) \mathrm{d} x_{i} ,\; \forall i,
\end{equation}
where $\mathscr{P}(x_{i}|\mathbf{H},\widetilde{\mathbf{y}})$ is the
marginal posterior distribution of the joint posterior distribution
$\mathscr{P}(\mathbf{x}|\mathbf{H},\widetilde{\mathbf{y}})$.
According to Bayes' rule, the joint posterior distribution obeys
\begin{equation}\label{eq:joint_def}
      \mathscr{P}(\mathbf{x}|\mathbf{H},\widetilde{\mathbf{y}}) \propto \mathscr{P}(\widetilde{\mathbf{y}}|\mathbf{H},\mathbf{x}) \mathscr{P}_{\mathsf{o}}(\mathbf{x}),
\end{equation}
where $\mathscr{P}(\widetilde{\mathbf{y}}|{\bf H},\mathbf{x})$ is
the likelihood function, $\mathscr{P}_{\mathsf{o}}(\mathbf{x})$ is
the prior distribution of the system state $\mathbf{x}$, and
$\propto$ denotes that the distribution is to be normalized so that
it has a unit integral.\footnote{On the basis of Bayes' theorem,
(\ref{eq:joint_def}) is originally expressed as
\begin{equation}\label{eq:joint_def11}
 \mathscr{P}(\mathbf{x}|\mathbf{H},\widetilde{\mathbf{y}}) = \frac{  \mathscr{P}(\widetilde{\mathbf{y}}|\mathbf{H},\mathbf{x}) \mathscr{P}_{\mathsf{o}}(\mathbf{x}) }
 {\mathscr{P}(\widetilde{\mathbf{y}}|\mathbf{H}) },
\end{equation}
where the denominator
\begin{equation}\label{eq:joint_def22}
 \mathscr{P}(\widetilde{\mathbf{y}}|\mathbf{H}) = \int  \mathscr{P}(\widetilde{\mathbf{y}}|\mathbf{H},\mathbf{x}) \mathscr{P}_{\mathsf{o}}(\mathbf{x}) \mathrm{d}\mathbf{x}
\end{equation}
defines the ``prior predictive distribution'' of
$\widetilde{\mathbf{y}}$ for a given topology matrix $\mathbf{H}$
and may be set to an unknown \emph{constant}. In calculating the
density of $\mathbf{x}$, any function that does not depend on this
parameter, such as $\mathscr{P}(\widetilde{\mathbf{y}}|\mathbf{H})$,
can be discarded. Therefore, by removing
$\mathscr{P}(\widetilde{\mathbf{y}}|\mathbf{H})$ from
(\ref{eq:joint_def11}), the relationship changes from being
``equals'' to being ``proportional.'' That is,
$\mathscr{P}(\mathbf{x}|\mathbf{H},\widetilde{\mathbf{y}})$ is
proportional to the numerator of (\ref{eq:joint_def11}). However, in
discarding $\mathscr{P}(\widetilde{\mathbf{y}}|\mathbf{H})$, the
density $\mathscr{P}(\mathbf{x}|\mathbf{H},\widetilde{\mathbf{y}})$
has lost some properties,  such as integration to one over the
domain of $\mathbf{x}$. To ensure that
$\mathscr{P}(\mathbf{x}|\mathbf{H},\widetilde{\mathbf{y}})$ is
properly distributed, the symbol $\propto$ simply means that the
distribution should be normalized to have a unit integral.}

Given that the entries of the measurement noise vector ${\bf e}$ are i.i.d. random variables and under the assumption that the prior distribution of $\mathbf{x}$ has a \emph{separable} form, that is, $\mathscr{P}_{\mathsf{o}}(\mathbf{x}) = \prod_{i= 1}^{N} \mathscr{P}_{\mathsf{o}}(x_{i})$, (\ref{eq:joint_def}) can be further factored as
\begin{equation}\label{eq:joint}
      \mathscr{P}(\mathbf{x}|\mathbf{H},\widetilde{\mathbf{y}}) \propto \prod_{\mu = 1}^{P} \mathscr{P}(\widetilde{y}_{\mu}|\mathbf{H},\mathbf{x})\prod_{i= 1}^{N} \mathscr{P}_{\mathsf{o}}(x_{i}),
\end{equation}
where $\mathscr{P}_{\mathsf{o}}(x_{i})$ is the prior distribution of
the $i$-th element of $\mathbf{x}$ and
$\mathscr{P}(\widetilde{y}_{\mu}|\mathbf{H},\mathbf{x})$ describes
the $\mu$-th measurement with i.i.d. complex Gaussian noise
\cite{Wen-15}, which can be explicitly represented as follows:
\begin{equation}
\mathscr{P}(\widetilde{y}_{\mu}|\mathbf{H},\mathbf{x}) = \int_{\widetilde{y}_{\mu}-\frac{\Delta}{2}}^{\widetilde{y}_{\mu}+\frac{\Delta}{2}} \mathscr{N}_{\mathbb{C}}\left( y_{\mu}; \sum_{i=1}^{N} H_{\mu i} x_{i}, \sigma^2\right) {\rm d} y_{\mu},
\end{equation}
where $H_{\mu i}$ denotes the component of $\mathbf{H}$ in the
$\mu$-th row and $i$-th column. For the considered problem, the
entries of the system state $\mathbf{x}$ can be treated as i.i.d.
complex Gaussian random variables with mean $\nu_{x}$ and variance
$\sigma^2_{x}$ for each prior distribution
$\mathscr{P}_{\mathsf{o}}(x_{i})$, that is,
$\mathscr{P}_{\mathsf{o}}(x_{i})=\mathscr{N}_{\mathbb{C}}(\nu_{x},\sigma^2_{x})$
\cite{Hu-11-CIM}. For brevity, the prior distribution of $x_{i}$ is
characterized by the prior parameter
$\boldsymbol{\theta}_{\mathsf{o}}=\{\nu_{x},\sigma^2_{x}\}$.

\begin{figure}
\begin{center}
\resizebox{3.5in}{!}{%
\includegraphics*{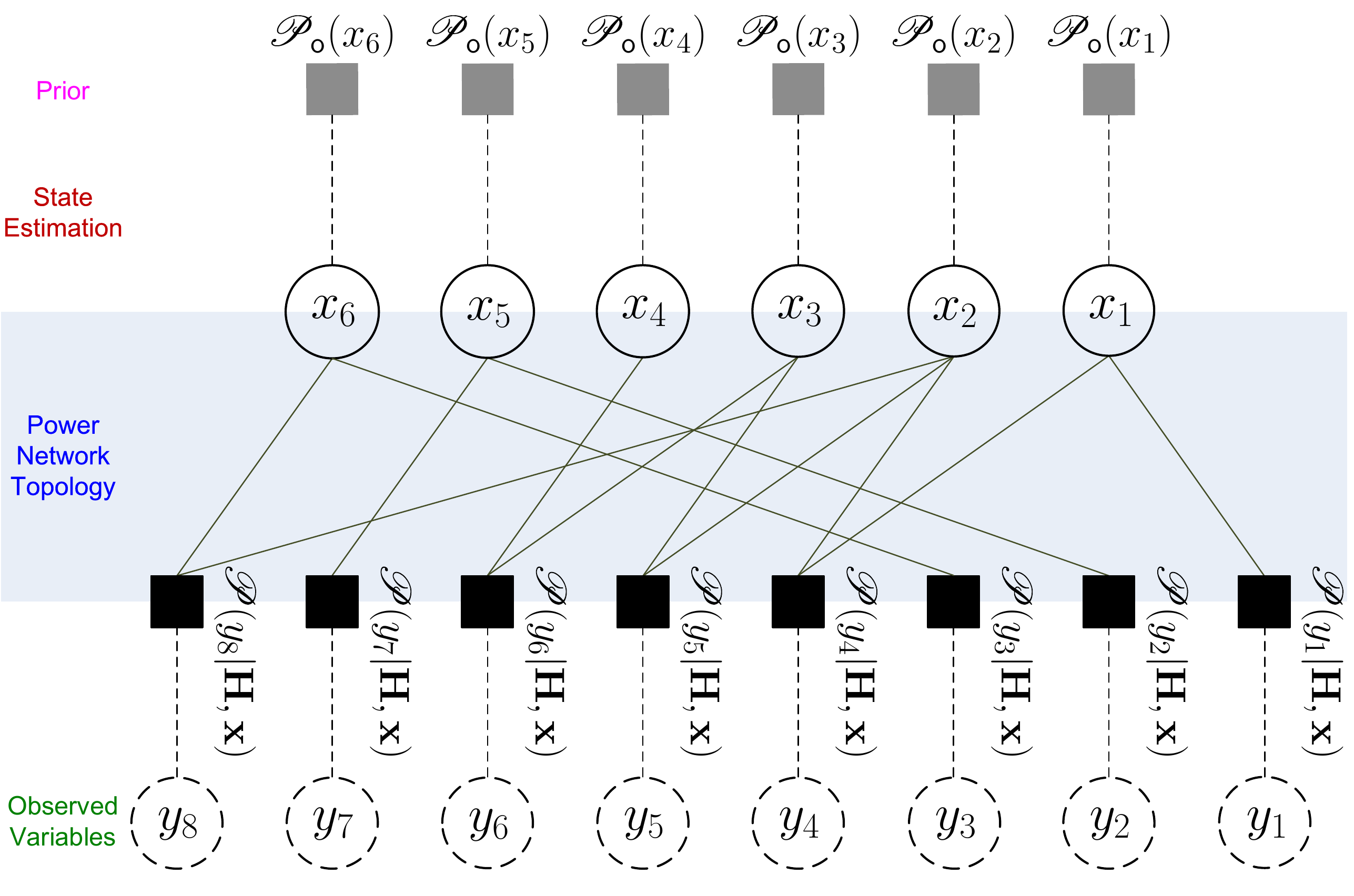} }%
\caption{Factor graph representing the state estimation problem for Fig. \ref{fig:system_model_example}.
The unobserved/observed random variables are depicted as solid/dashed lines with open circles, and the factor nodes as depicted as black squares.
The factor nodes ensure (in probability) the condition $\widetilde{y}_{\mu} = {\cal Q}_{\mathbb{C}}\left( \sum_{i} H_{\mu i} x_{i} + e_{\mu} \right)$.}\label{fig:factorgraph_example}
\end{center}
\end{figure}

The decomposition of the joint distribution in (\ref{eq:joint}) can be well represented by a factor graph $\mathcal{G}=(\mathcal{V},\mathcal{F},\mathcal{E})$, where $\mathcal{V}=\{x_i\}_{i=1}^{N}$ is the set of \emph{unobserved} variable nodes, $\mathcal{F}=\{\mathscr{P}(\widetilde{y}_{\mu}|\mathbf{H},\mathbf{x})\}_{\mu=1}^{P}$ is the set of factor nodes, where each factor node ensures (in probability) the condition $\widetilde{y}_{\mu} = {\cal Q}_{\mathbb{C}}\left( \sum_{i} H_{\mu i} x_{i} + e_{\mu} \right)$, and $\mathcal{E}$ denotes the set of edges.
Specifically, edges indicate the involvement between function nodes and variable nodes; that is, an edge between variable node $x_i$ and factor node $\mathscr{P}(\widetilde{y}_{\mu}|\mathbf{H},\mathbf{x})$ indicates that the given factor function $\mathscr{P}(\widetilde{y}_{\mu}|\mathbf{H},\mathbf{x})$ is a function of $x_i$. Fig. \ref{fig:factorgraph_example} provides a factor graph representation for the fictitious six buses system shown in Fig. \ref{fig:system_model_example}.

Given the factor graph representation, message-passing-based algorithms, such as \emph{belief propagation} (BP) \cite{Pea-book-88,FG-01-IT}, can be applied to compute $\mathscr{P}(x_{i}|\mathbf{H},\widetilde{\mathbf{y}})$
approximately. Methodically, BP passes the following ``messages,'' which denote the probability distribution functions, along the edges of the graph as follows \cite{MP-book}:
\begin{align}
    \hspace{-0.35cm}  \mathscr{M}_{i \rightarrow \mu}^{(t+1)}(x_{i})  & \propto \mathscr{P}_{\mathsf{o}}(x_{i}) \prod_{\begin{subarray}{c} \gamma \,\in\, \mathscr{L}(i)\\ \gamma \,\neq\, \mu \end{subarray}} \mathscr{M}_{\gamma \rightarrow i}^{(t)}(x_{i}), \label{m_f_c_v} \\
    \hspace{-0.35cm}  \mathscr{M}_{\mu \rightarrow i}^{(t+1)}(x_{i})  & \propto \!\!\int\!\!\!\!\! \prod_{\begin{subarray}{c} k \,\in\, \mathscr{L}(\mu)\\ k \,\neq\, i \end{subarray}}\!\!\!\!\!\! \mathrm{d}x_{k} \! \Bigg[\!\mathscr{P}(\widetilde{y}_{\mu}|\mathbf{H},\mathbf{x})\!\!\!\!\!   \prod_{\begin{subarray}{c} k \,\in\, \mathscr{L}(\mu)\\ k \,\neq\, i \end{subarray}}\!\!\!\!\!\! \mathscr{M}_{k \rightarrow \mu}^{(t)}(x_{k})\Bigg],\!\! \!\label{m_v_2_f}
\end{align}
where superscript $t$ indicates the iteration number, $\mathscr{M}_{i \rightarrow \mu}(x_{i}) $ is the message from the $i$-th variable node to the $\mu$-th factor node, $\mathscr{M}_{\mu \rightarrow i}(x_{i})$ is the message from the $\mu$-th factor node to the $i$-th variable node,  $\mathscr{L}(i)$ is the set of factor nodes that are neighbors of the $i$-th variable node, and $\mathscr{L}(\mu)$ is the set of variable nodes that are neighbors of the $\mu$-th factor node.
Then, the approximate marginal distribution is computed according to the following equation:
\begin{equation}\label{eq:marginal_App}
      \mathscr{P}^{(t)}(x_{i}|\mathbf{H},\mathbf{y}) \propto \mathscr{P}_{\mathsf{o}}(x_{i}) \prod_{\mu \,\in\, \mathscr{L}(i)} \mathscr{M}_{\mu \rightarrow i}^{(t)}(x_{i}).
\end{equation}

\begin{algorithm}[tbp]\label{ago:ago1}  \footnotesize
  \caption{{\tt EMSwGAMP} algorithm}
  \KwIn{$\widetilde{\mathbf{y}}, \mathbf{H}, \sigma$}
  \KwOut{$\{\widehat{x}_{i}^{(t)}\}_{i=1}^{N}$}
  \SetKwFunction{KwFn}{Fn}
  $\widehat{\mathbf{x}}^{\left(0\right)}=\{\widehat{x}_{i}^{\left(0\right)}\}_{i=1}^{N} \leftarrow \{ 1 \}$\\
  $\boldsymbol{\tau}^{\left(0\right)}=\{\tau_{i}^{\left(0\right)}\}_{i=1}^{N} \leftarrow \{1\}$\\
  $\boldsymbol{\varrho}^{\left(0\right)}=\{{\varrho}_{\mu}^{\left(0\right)}\}_{\mu = 1}^{P} \leftarrow \{1\}$\\
  $\boldsymbol{\omega}^{\left(0\right)}=\{{\omega}_{\mu}^{\left(0\right)}\}_{\mu = 1}^{P} \leftarrow \{\widetilde{y}_{\mu}\}$\\
  $t \leftarrow 1$ \\
  \While{Stopping criteria are not met}{
  \For{$\mu=1$ \emph{\KwTo} $P$}{
      $\varrho^{(t)}_{\mu}           \leftarrow \sum_{i=1}^{N} \left| H_{\mu i}\right|^{2} \tau_{i}^{(t-1)}$ \\
      $\omega_{\mu}^{(t)}      \leftarrow \sum_{i=1}^{N} H_{\mu i} \widehat{x}_{i}^{(t-1)} - \frac{\widetilde{y}_{\mu} - \omega_{\mu}^{(t-1)}}{\sigma+\varrho^{(t-1)}_{\mu}}\varrho^{(t)}_{\mu}$ \\
      $\varsigma_{\mu}^{(t)}   \leftarrow \mathbb{VAR} \left[z_{\mu}|\widetilde{y}_{\mu}, \omega_{\mu}^{(t)}, \varrho_{\mu}^{(t)}\right]$\\
      $\widehat{z}_{\mu}^{(t)} \leftarrow \mathbb{E}\left[z_{\mu}|\widetilde{y}_{\mu}, \omega_{\mu}^{(t)}, \varrho_{\mu}^{(t)}\right]$\\
      $\widehat{s}_{\mu}^{(t)} \leftarrow \frac{\widehat{z}_{\mu}^{(t)} - \omega_{\mu}^{(t)}}{\varrho_{\mu}^{(t)}} $\\
      $\zeta_{\mu}^{(t)}       \leftarrow \left(1- \frac{\varsigma_{\mu}^{(t)}}{\varrho_{\mu}^{(t)}}\right)\frac{1}{\varrho_{\mu}^{(t)}}$ \\

      }
      $[\ell_1, \ell_2, \cdots, \ell_{N} ] \leftarrow {\tt permute}(\{1,2,\cdots,N\})$ \\
      \For{$k=1$ \emph{\KwTo} $N$}{
      $i\leftarrow \ell_k$\\
      $(\Sigma_{i}^{(t)})^2      \leftarrow \left[\sum_{\mu =1}^{P} \left|H_{\mu i}\right|^{2} \zeta_{\mu}^{(t)} \right]^{-1}$\\
      $R_{i}^{(t)}               \leftarrow \widehat{x}_{i}^{(t-1)} +  (\Sigma_{i}^{(t)})^2 \sum_{\mu =1}^{P} H_{\mu i}^{*}\, \widehat{s}_{\mu}^{(t)}$\\
      $\widehat{x}_{i}^{(t)}     \leftarrow \mathbb{E}{\left[ x_{i} | \boldsymbol{\theta}_{\mathsf{o}}, R_{i}^{(t)}, (\Sigma_{i}^{(t)})^2 \right]} $\\
      $\tau_{i}^{(t)}            \leftarrow \mathbb{VAR}{\left[ x_{i} | \boldsymbol{\theta}_{\mathsf{o}}, R_{i}^{(t)}, (\Sigma_{i}^{(t)})^2 \right]} $ \\
      \For{$\mu \in \mathscr{L}(i)$ }{
      $\breve{\varrho}_{\mu}^{(t)} \leftarrow \varrho_{\mu}^{(t)}$\\
      $\varrho_{\mu}^{(t)} \leftarrow \breve{\varrho}_{\mu}^{(t)} + \left|H_{\mu i}\right|^{2} (\tau_{i}^{(t)} - \tau_{i}^{(t-1)} )$\\
      $\omega_{\mu}^{(t)}     \leftarrow \omega_{\mu}^{(t)} + H_{\mu i} (\widehat{x}_{i}^{(t)} - \widehat{x}_{i}^{(t-1)}) - \frac{\widetilde{y}_{\mu} - \omega_{\mu}^{(t-1)}}{\sigma+\varrho^{(t-1)}_{\mu}}(\varrho^{(t)}_{\mu}- \breve{\varrho}_{\mu}^{(t)})$
      }

      }
      $\nu_{x} \leftarrow \frac{1}{N}\sum_{i=1}^{N} R_{i}^{(t)}$\\
      $\sigma_{x}^{2} \leftarrow \frac{1}{N}\sum_{i=1}^{N} \left[(\nu_{x}- R_{i}^{(t)})^2+(\Sigma_{i}^{(t)})^2\right]$\\

  $t \leftarrow t+1$
  }

\end{algorithm}

\subsection{EM-assisted GAMP for the State Estimation Problems}

However, according to \cite{Guo-06-ITW,Wang-14-WCNC,Rangan-11-ISIT}, BP remains \emph{computationally intractable} for large-scale problems because of the high-dimensional integrals involved and large number of messages required.
Moreover, the prior parameters $\boldsymbol{\theta}_{\mathsf{o}}$ and $\sigma^2$ are usually \emph{unknown} in advance.
Fortunately, BP can be simplified as the GAMP algorithm \cite{Rangan-11-ISIT} based on the central limit theorem and Taylor expansions to enhance computational tractability.\footnote{To state this more precisely, applying the central limit theorem to messages yields a Gaussian approximation that can be used to simplify BP from a recursion on functions to a recursion on numbers. On the basis of a series of Taylor expansions,
the number of messages can be reduced significantly.}
By contrast, the EM algorithm can be applied to learn the prior parameters \cite{Vila-13TSP}. With the aid of these two algorithms, we develop an iterative method involving the following two phases per iteration for the state estimation problems:
First, ``swept'' GAMP ({\tt SwGAMP}) \cite{SwAMP-2015}, a modified version of GAMP, is exploited to estimate $\mathbf{x}$.
Second, the EM algorithm is applied to learn the prior parameters from the data at hand.
The stepwise implementation procedure of the proposed state estimation scheme, referred to as {\tt EMSwGAMP}, is presented in Algorithm \ref{ago:ago1}.
Considering the space limitations, a detailed derivation of (complex) GAMP and EM is excluded in this paper.
For the details, refer to \cite{Rangan-11-ISIT,Krzakala-12,SwAMP-2015,Jay-15-TVT,Vila-13TSP}.
Then, we provide several explanations for each line of Algorithm \ref{ago:ago1} to ensure better understanding of the proposed scheme.

In Algorithm \ref{ago:ago1}, $\widehat{x}_{i}^{(t)}$ denotes the
estimate of the $i$-th element of $\mathbf{x}$ in the $t$-th
iteration and $\tau_{i}^{(t)}$ can be interpreted as an
approximation of the posterior variance of $\widehat{x}_{i}^{(t)}$;
these two quantities are initialized as $\widehat{x}_{i}^{(0)} = 1$
and $\tau_{i}^{(0)}=1$, respectively. For each factor node, we
introduce two auxiliary variables $\omega_{\mu}^{(t)}$ and
$\varrho_{\mu}^{(t)}$, given in Lines 9 and 8 of Algorithm
\ref{ago:ago1}, describing the current mean and variance estimates
of the $\mu$-th element of $\widetilde{\mathbf{y}}$, respectively.
The initial conditions of $\omega_{\mu}^{(0)}$ and
$\varrho_{\mu}^{(0)}$ are specified in Lines 4 and 3 of Algorithm
\ref{ago:ago1}, respectively. As stated previously, $z_{\mu}$ is the
$\mu$-th element of the \emph{noise free} measurement vector
$\mathbf{z}$, that is, $z_{\mu} = \sum_{i} H_{\mu i}\,x_{i}$.
Therefore, according to the derivations of \cite{Rangan-11-ISIT},
$z_{\mu}$ conditioned on $x_i$ can be further approximated as a
Gaussian distribution with the mean and variance given in Lines 9
and 8 of Algorithm \ref{ago:ago1}, respectively, which are evaluated
with respect to the following expression
\begin{equation} \label{eq:post_z}
\mathscr{P}(z_{\mu}|\widetilde{y}_{\mu}) \propto \int_{\widetilde{y}_{\mu}-\frac{\Delta}{2}}^{\widetilde{y}_{\mu}+\frac{\Delta}{2}} \mathscr{N}_{\mathbb{C}}(y_{\mu}; z_{\mu}, \sigma^2)\mathscr{N}_{\mathbb{C}}(z_{\mu}; \omega_{\mu}^{(t)}, \varrho_{\mu}^{(t)}) {\rm d} y_{\mu}.
\end{equation}
(\ref{eq:post_z}) represents the quantization noise that can be
regarded as a Gaussian distribution whose mean is $\omega_{\mu}$ and
variance is $\varrho_{\mu}$. Finally, the messages from factor nodes
to variable nodes are reduced to a simple message, which is
parameterized by $\widehat{s}_{\mu}$ and $\zeta_{\mu}$, given in
Lines 12 and 13 of Algorithm \ref{ago:ago1}. Therefore, we refer to
messages $\{\widehat{s}_{\mu},\zeta_{\mu}\}$ as measurement updates.

Similarly, for the variable nodes, we also introduce two auxiliary
variables $R_{i}^{(t)}$ and $(\Sigma_{i}^{(t)})^2$, given in Lines
18 and 17 of Algorithm \ref{ago:ago1}, describing the current mean
and variance estimates of the $i$-th element of $\mathbf{x}$
\emph{without} considering the prior information of $x_{i}$,
respectively. Then, adding the prior information of $x_{i}$, that
is,
$\mathscr{P}_{\mathsf{o}}(x_{i})=\mathscr{N}_{\mathbb{C}}(\nu_{x},\sigma^2_{x})$,
to the message updates, the posterior mean and variance of $x_{i}$
are given in Lines 19 and 20 of Algorithm \ref{ago:ago1},
respectively where is considered with respect to the following
expression:
\begin{equation} \label{eq:post_x}
\!    \mathscr{P}(x_{i};\widehat{x}_{i}^{(t)},\tau_{i}^{(t)})\! \propto\! \mathscr{N}_{\mathbb{C}}(x_i ; \nu_{x}, \sigma^2_{x}) \mathscr{N}_{\mathbb{C}}(x_i; R_{i}^{(t)}, (\Sigma_{i}^{(t)})^2).\!
\end{equation}
Here, $\widehat{x}_{i}^{(t)}$ (i.e., the calculation of Expectation
in Line 19 of Algorithm 1) and $\tau_{i}^{(t)}$ (i.e., the
calculation of VAR in Line 20 of Algorithm 1) can be easily obtained
using  the standard formulas for Gaussian distributions as
\cite{Wang-15-ICC,Barbier-15}
\begin{align}
 \widehat{x}_{i}^{(t)} & = R_{i}^{(t)} + \frac{(\Sigma_{i}^{(t)})^2 }{(\Sigma_{i}^{(t)})^2 + \sigma^2_{x}} {\left(\nu_{x} - R_{i}^{(t)} \right)},\\
        \tau_{i}^{(t)}  & = \frac{(\Sigma_{i}^{(t)})^2 \sigma^2_{x}}{(\Sigma_{i}^{(t)})^2 + \sigma^2_{x}}. \label{eq:def_fc}
\end{align}

Manoel et al. \cite{SwAMP-2015} slightly modified the update scheme for GAMP, where partial quantities are updated \emph{sequentially} rather than \emph{in parallel}, to improve the stability of GAMP.
\footnote{The empirical studies demonstrate that GAMP with slight modifications not only exhibits good convergence performances but is also more robust to difficult measurement matrix ${\bf H}$ as compared with the original GAMP.}
Specifically, $\sum_{i}  H_{\mu i} \widehat{x}_{i}^{(t-1)}$ and $\varrho_{\mu}^{(t)}$ are recomputed as the sweep updates over $i$ for a single iteration step.
Lines 22-24 of Algorithm \ref{ago:ago1} are the core steps to perform the so-called sweep (or reordering) updates.
In brief, we refer to messages $\{\widehat{x}_{i},\tau_{i}\}$ as variable updates and to messages $\{\widehat{s}_{\mu},\zeta_{\mu}\}$ as measurement updates for the {\tt SwGAMP} algorithm. One iteration of the {\tt SwGAMP} algorithm involves the implementation of these updates together with the estimation of the system state $\mathbf{x}$.

In the first phase of Algorithm \ref{ago:ago1}, the prior parameters $\boldsymbol{\theta}_{\mathsf{o}}= \{\nu_{x},\sigma^2_{x}\}$ are treated as known parameters, but may be unknown in practice.
Thus, the second phase of the proposed algorithm is to adopt the EM algorithm to learn the prior parameters $\boldsymbol{\theta}_{\mathsf{o}}$ on the basis of the quantities acquired in the first phase of the algorithm.
The EM algorithm is a general iterative method for likelihood optimization in probabilistic models with hidden variables.
In our case, the EM-updates will be expressed in the following form \cite{Vila-13TSP}
\begin{equation} \label{eq:EM_update}
     \boldsymbol{\theta}_{\mathsf{o}}^{\rm new} = \argmax_{\boldsymbol{\theta}_{\mathsf{o}}} \mathbb{E}\left\{ \ln \mathscr{P}(\mathbf{x},\mathbf{y};\boldsymbol{\theta}_{\mathsf{o}}) \right\},
\end{equation}
where the expectation takes over the posterior probability of
$\mathbf{x}$ conditioned on $\boldsymbol{\theta}_{\mathsf{o}}$ and
$\mathbf{y}$. Following similar steps in \cite{Vila-13TSP},  we can
derive a set of EM-based update equations for the hyperparameters,
that is, the \emph{prior} information of the system states (i.e.,
$\nu_{x}$ and $\sigma^2_{x}$) that should be inferred. The detailed
EM updates for the hyperparameters are provided in Lines 25 and 26
of Algorithm \ref{ago:ago1}, respectively. Notably, the quantities
$\{R_{i}^{(t)}\}_{i=1}^{N}$, $\{(\Sigma_{i}^{(t)})^2\}_{i=1}^{N}$,
$\{\widehat{z}_{\mu}^{(t)} \}_{\mu=1}^{P}$, and
$\{\varsigma_{\mu}^{(t)}\}_{\mu = 1}^{P}$ are readily available
after running the {\tt SwGAMP} algorithm in the first phase.

{\it Remark 3.1 (Calculating Lines 10 and 11 of Algorithm
\ref{ago:ago1} with high resolution representation of the measured
data) :} In modern SCADA systems, each measurement is quantized and
represented using a word length of 12 (or 16) bits. With such high
precision representation of the measurement data, the error between
the quantized value $\widetilde{y}_{\mu}$ and the actual value
$y_{\mu}$ can be negligible, that is, $\widetilde{y}_{\mu}
\simeq{y}_{\mu}$. In this case, (\ref{eq:post_z}) can be rewritten
as follows:
\begin{equation} \label{eq:post_z_unq}
\mathscr{P}(z_{\mu}|\widetilde{y}_{\mu})\propto \mathscr{N}_{\mathbb{C}}(y_{\mu};
z_{\mu}, \sigma^2) \mathscr{N}_{\mathbb{C}}(z_{\mu};
\omega_{\mu}^{(t)}, \varrho_{\mu}^{(t)}).
\end{equation}
Then, the moments, $\widehat{z}_{\mu}^{(t)}$ and $\varsigma_{\mu}^{(t)}$, can be easily obtained using standard formulas for Gaussian distributions, as follows \cite{Wang-15-ICC}:
\begin{align}
      \widehat{z}_{\mu}^{(t)}  &
      = \omega_{\mu}^{(t)} + \frac{ \varrho_{\mu}^{(t)}}{\varrho_{\mu}^{(t)} + \sigma^2} {\left(\widetilde{y}_{\mu} -  \omega_{\mu}^{(t)}\right)}, \label{z_mean} \\
      \varsigma_{\mu}^{(t)}    &
      = \frac{ \varrho_{\mu}^{(t)} \sigma^2}{\varrho_{\mu}^{(t)} + \sigma^2}. \label{z_var} \vspace{-.05in}
\end{align}

{\it Remark 3.2 (Calculating Lines 10 and 11 of Algorithm
\ref{ago:ago1} under the ``quantized'' scenario) :} When
quantization error is \emph{nonnegligible}, particularly at coarse
quantization levels, (\ref{eq:post_z_unq}) is no longer valid
because of the fact that using $\widetilde{y}_{\mu}$ to approximate
${y}_{\mu}$ will result in severe performance degradation. In this
case, we have to adopt (\ref{eq:post_z}) to determine the
conditional mean $\widehat{z}_{\mu}^{(t)}$ and conditional variance
$\varsigma_{\mu}^{(t)}$, which can be obtained as follows:
\begin{align}
      \widehat{z}_{\mu}^{(t)}  &
      = \frac{ \int_{\widetilde{y}_{\mu}-\frac{\Delta}{2}}^{\widetilde{y}_{\mu}+\frac{\Delta}{2}} y_{\mu} \mathscr{N}_{\mathbb{C}}(y_{\mu};\omega_{\mu}^{(t)}, \sigma^2+\varrho_{\mu}^{(t)}) {\rm d} y_{\mu} }
      { \int_{\widetilde{y}_{\mu}-\frac{\Delta}{2}}^{\widetilde{y}_{\mu}+\frac{\Delta}{2}} \mathscr{N}_{\mathbb{C}}(y_{\mu};\omega_{\mu}^{(t)}, \sigma^2+\varrho_{\mu}^{(t)}) {\rm d} y_{\mu} }, \label{Q_z_mean} \\
      \varsigma_{\mu}^{(t)}    &
      = \frac{ \int_{\widetilde{y}_{\mu}-\frac{\Delta}{2}}^{\widetilde{y}_{\mu}+\frac{\Delta}{2}} | y_{\mu} - \widehat{y}_{\mu}^{(t)} |^2 \mathscr{N}_{\mathbb{C}}(y_{\mu};\omega_{\mu}^{(t)}, \sigma^2+\varrho_{\mu}^{(t)}) {\rm d} y_{\mu} }
      { \int_{\widetilde{y}_{\mu}-\frac{\Delta}{2}}^{\widetilde{y}_{\mu}+\frac{\Delta}{2}} \mathscr{N}_{\mathbb{C}}(y_{\mu};\omega_{\mu}^{(t)}, \sigma^2+\varrho_{\mu}^{(t)}) {\rm d} y_{\mu} }. \label{Q_z_var}
\end{align}
Explicit expressions of (\ref{Q_z_mean}) and (\ref{Q_z_var}) are provided in \cite{Wen-15}.

{\it Remark 3.3 (Stopping criteria):} The algorithm can be
discontinued either when a predefined number of iterations is
reached or when it converges in the relative difference of the norm
of the estimate of $\mathbf{x}$, or both. The relative difference of
the norm is given by the quantity $\epsilon \triangleq \sum_{i}^{N}
|\widehat{x}_{i}^{(t)} - \widehat{x}_{i}^{(t-1)}|^2$.

\section{Simulation Results and Discussion}\label{sec:04}

In this section, we evaluate the performance of the proposed {\tt
EMSwGAMP} algorithm for single-phase state and three-phase state
estimations. The optimal PMU placement issue is \emph{not} included
in this study, and we assume that PMUs are placed in terminal buses.
In the single-phase state estimation, IEEE 69-bus radial
distribution network \cite{69-bus-data} is used for the test system,
where the subset of buses with PMU measurements is denoted by
$\mathcal{P}_{69}=\{1, 27, 35, 46, 50, 52, 67, 69\}$. A
\emph{modified} version of IEEE $69$-bus radial distribution
network, referred to as 69m in this study, is examined to verify the
robustness of the proposed algorithm. The system settings of this
modified test system are identical to those of the IEEE 69-bus
radial distribution network, with the exception of the bus voltages
of this test system being able to vary within a large range, thereby
increasing the load levels of this system. For these two test
system, we have $68$ current measurements and $8$ voltage
measurements. The software toolbox MATPOWER \cite{MATPOWER} is
utilized to run the proposed state estimation algorithm for various
cases in the single-phase state estimation. The IEEE 37-bus
three-phase system is used as the test system for the three-phase
state estimation, where the subset of buses with PMU measurements is
denoted by $\mathcal{P}_{37}=\{1, 10, 15, 20, 29, 35\}$. In contrast
to the single-phase state estimation, the system state of the
three-phase estimation is generated by test system documents instead
of MATPOWER. We have $105$ current measurements and $18$ voltage
measurements in 37-bus three-phase system. Prior distributions of
the voltage at each bus for different test systems can be found in
Table \ref{tb:initial_priori}. In each estimation, the mean squared
error (MSE) of the bus voltage magnitude and that of the bus voltage
phase angle are used as comparison measures, which are expressed as
${\rm MSE} = \frac{1}{N}\sum_{i=1}^{N}|x_{i}-\widehat{x}_{i}|^2$,
${\rm MSE}_{\rm magn} =
\frac{1}{N}\sum_{i=1}^{N}(|x_{i}|-|\widehat{x}_{i}|)^2$, and ${\rm
MSE}_{\rm phase} =
\frac{1}{N}\sum_{i=1}^{N}[\arg(x_{i})-\arg(\widehat{x}_{i})]^2$,
respectively. The LMMSE estimator is tested for comparison. In our
implementation, termination of Algorithm \ref{ago:ago1} is declared
when the corresponding constraint violation is less than $\epsilon
=10^{-8}$. A total of $1,000$ Monte Carlo simulations were conducted
and evaluated to obtain average results and to analyze the achieved
measures. The simulations for computation time were conducted
utilizing an Intel i7-4790 computer with 3.6 GHz CPU and 16 GB RAM.
For clarity, the number of measurements quantized to be
$\mathcal{B}$-bit is denoted as $\mathcal{K}$, where $\mathcal{B}$
denotes the number of bits used for quantization.

\begin{table}
\begin{center}
\caption{The prior distribution of bus voltage for different
systems}\label{tb:initial_priori}
\begin{tabular}{|C{1.35cm}|c|c|c|c|c|c|c|}
  \hline
                                 & \multirow{2}{*}{$N$} & Magnitude   & Phase                    &\multirow{2}{*}{$\sigma^2_{x}$} \\
                                 &                      & (mean)      & (mean)                   &  \\   \hline\hline
 \multirow{2}{*}{single-phase}   &\multirow{1}{*}{69\phantom{m}}   & $1.00$      & $5.60 \times  10^{-4}$   & $5.46 \times10^{-7}$  \\    \cline{2-5}
                                 &\multirow{1}{*}{69m}             & $1.04$      & $1.71 \times 10^{-2}$   & $5.66 \times 10^{-4}$  \\    \hline
 three-phase                     &\multirow{1}{*}{37\phantom{m}}   & $0.01$      & $1.12$   & $9.86 \times 10^{-1}$  \\    \hline
\end{tabular}
\end{center}
\end{table}

\begin{table}
\begin{center}
\caption{Average estimation results obtained by {\tt EMSwGAMP} and LMMSE with the unquantized measured data for the various
systems}\label{tb:gaussian_model_test_result}
\begin{tabular}{|c|c|c|c|c|c|}
   \hline
   \multirow{2}{*}{$N$} & \multirow{2}{*}{Algorithm} & \multirow{2}{*}{${\rm MSE}$} & \multirow{2}{*}{${\rm MSE}_{\rm magn}$} & \multirow{2}{*}{${\rm MSE}_{\rm phase}$} \\
     &&& &  \\  \hline\hline
   \multirow{2}{*}{69\phantom{m}}  & {\tt EMSwGAMP} & $3.84 \times 10^{-4}$ &  $2.06 \times 10^{-4}$ & $7.78 \times 10^{-4}$   \\  \cline{2-5}
                        & LMMSE          & $8.16 \times 10^{-4}$ &  $4.65 \times 10^{-4}$  &  $3.55 \times 10^{-4}$   \\  \hline \hline
   \multirow{2}{*}{69m} & {\tt EMSwGAMP} & $7.11 \times 10^{-4}$ &  $3.00 \times 10^{-4}$  &  $3.78 \times 10^{-4}$   \\  \cline{2-5}
                        & LMMSE          & $8.22 \times 10^{-4}$ &  $4.75 \times 10^{-4}$  &  $3.23 \times 10^{-4}$   \\  \hline
\end{tabular}
\end{center}
\end{table}

Table \ref{tb:gaussian_model_test_result} shows a summary of the
average ${\rm MSE}$, ${\rm MSE}_{\rm magn}$, and ${\rm MSE}_{\rm
phase}$ achieved by {\tt EMSwGAMP} and LMMSE for single-phase state
estimation with various systems. The results show that even under
the traditional \emph{unquantized} setting,\footnote{As mentioned in
Remark 3.1, when the measured data are represented using a
wordlength of 16 bits, the quantization error can be negligible. In
this case, such high-precision measurement data are henceforth
referred to as the \emph{unquantized} measured data. Therefore, all
measurements are quantized with $16$ bits in In Table
\ref{tb:gaussian_model_test_result}.} {\tt EMSwGAMP} still
outperforms LMMSE because {\tt EMSwGAMP} exploits the statistical
knowledge of the estimated parameters
$\boldsymbol{\theta}_{\mathsf{o}}$, which is learned from the data
via the second phase of {\tt EMSwGAMP}, that is, the EM learning
algorithm. Table \ref{tb:gaussian_model_parameter_learning} reveals
that the estimation results of the system states using {\tt
EMSwGAMP} are close to the true values, which validates the
effectiveness of the proposed learning algorithm. From a detailed
inspection of Table \ref{tb:gaussian_model_parameter_learning}, we
found that the mean value of voltage magnitude can be exactly
estimated through the EM learning algorithm. Therefore, the average
${\rm MSE}_{\rm magn}$ of {\tt EMSwGAMP} is better than that of
LMMSE. However, the mean value of voltage phase cannot be estimated
accurately by the EM learning algorithm. As a result, the average
${\rm MSE}_{\rm phase}$ of {\tt EMSwGAMP} is inferior to that of
LMMSE.

\begin{table}
\begin{center}
\caption{Parameter learning results using {\tt
EMSwGAMP}}\label{tb:gaussian_model_parameter_learning}
\begin{tabular}{|c|c|c|c|c|c|c|}
  \hline
    \multirow{2}{*}{$N$} &\multirow{2}{*}{Algorithm} & Magnitude   & Phase    \\
                         &                           & (mean)      & (mean)    \\   \hline\hline
   \multirow{2}{*}{69\phantom{m}}   & True value                & $1.00$   & $\phantom{-}5.60\times 10^{-4}$  \\    \cline{2-4}
                         & {\tt EMSwGAMP}            & $1.00$   & $-2.53 \times 10^{-4}$  \\    \hline \hline
    \multirow{2}{*}{69m} & True value                & $1.04$   & $\phantom{-}1.71 \times 10^{-2}$  \\    \cline{2-4}
                         & {\tt EMSwGAMP}            & $1.04$   & $\phantom{-}1.10 \times 10^{-2}$  \\    \hline
\end{tabular}
\end{center}
\end{table}

We consider an extreme scenario where several measurements are
quantized to ``$1$'' bit, but others are not, to reduce the amount
of transmitted data. The measurements selected to be quentized are
provided in Appendix A. Table \ref{tb:1bit_num_MSE_compare} shows
the average ${\rm MSE}$, ${\rm MSE}_{\rm magn}$, and ${\rm MSE}_{\rm
phase}$ against $\mathcal{K}$ obtained by {\tt EMSwGAMP}, where in
the performance of the LMMSE algorithm with $\mathcal{K}= 17$ is
also included for the purpose of comparison. The following
observations are noted on the basis of Table
\ref{tb:1bit_num_MSE_compare}: First, when the measurement is
quantized with 1 bit, we only know that the measurement is positive
or not so that the information related to the system is lost. Hence,
as expected, increasing $\mathcal{K}$ naturally degrades the average
MSE performance because more information is lost. However, the
achieved performance of the 69-bus test system is \emph{less}
sensitive to $\mathcal{K}$ because the bus voltage variations in
this system are small. Thus, the proposed algorithm can easily deal
with \emph{incomplete} data. Second, for the system with large bus
voltage fluctuations, the obtained ${\rm MSE}_{\rm phase}$
performance of the modified 69-bus test system can still achieve
$10^{-3}$ when $\mathcal{K}\leq 17$. However, with $\mathcal{K}=
17$, the LMMSE algorithm exhibits \emph{poor} performance for both
considered test systems, which cannot be used in practice.

\begin{table}
\begin{center}
\caption{Average estimation results obtained by {\tt EMSwGAMP} with different number of $1$-bit quantized measurements for the various systems}\label{tb:1bit_num_MSE_compare}
\begin{tabular}{|C{0.35cm}|C{1.25cm}|c|C{1.45cm}|C{1.45cm}|C{1.45cm}|}
  \hline
    \multirow{2}{*}{$N$} & \multirow{2}{*}{Algorithm} & \multirow{2}{*}{$\mathcal{K}$} & \multirow{2}{*}{${\rm MSE}$} & \multirow{2}{*}{${\rm MSE}_{\rm magn}$} & \multirow{2}{*}{${\rm MSE}_{\rm phase}$} \\
                         & &                                && &  \\  \hline\hline
  \multirow{10}{*}{69\phantom{m}} & \multirow{9}{*}{{\tt EMSwGAMP}}& $\phantom{0}0$  & $3.84 \times 10^{-4}$ &  $2.07 \times 10^{-4}$ & $1.77 \times 10^{-4}$ \\     \cline{3-6}
                      & & $\phantom{0}2$  & $1.00 \times 10^{-3}$ &  $5.97 \times 10^{-4}$ &  $4.44 \times 10^{-4}$ \\     \cline{3-6}
                      & & $\phantom{0}4$  & $1.00 \times 10^{-3}$ &  $6.18 \times 10^{-4}$ &  $4.74 \times 10^{-4}$ \\     \cline{3-6}
                      & & $17$            & $1.00 \times 10^{-3}$ &  $5.76 \times 10^{-4}$ &  $4.19 \times 10^{-4}$ \\     \cline{3-6}
                      & & $19$            & $1.00 \times 10^{-3}$ &  $5.72 \times 10^{-4}$ &  $4.34 \times 10^{-4}$ \\     \cline{3-6}
                      & & $23$            & $1.00 \times 10^{-3}$ &  $5.89 \times 10^{-4}$ &  $4.56 \times 10^{-4}$ \\     \cline{3-6}
                      & & $27$            & $1.10 \times 10^{-3}$ &  $6.04 \times 10^{-4}$ &  $4.87 \times 10^{-4}$ \\     \cline{3-6}
                      & & $34$            & $1.00 \times 10^{-3}$ &  $5.70 \times 10^{-4}$ &  $4.47 \times 10^{-4}$ \\     \cline{3-6}
                      & & $42$            & $1.00 \times 10^{-3}$ &  $5.59 \times 10^{-4}$ &  $4.41 \times 10^{-4}$ \\     \cline{2-6}
                & LMMSE & $17$            & $2.39 \times 10^{-1}$ &  $1.07 \times 10^{-1}$ &  $1.59 \times 10^{-1}$\\  \hline \hline
  \multirow{10}{*}{69m}& \multirow{9}{*}{{\tt EMSwGAMP}}& $\phantom{0}0$  & $7.00 \times 10^{-4}$ &  $3.90 \times 10^{-4}$ & $3.58 \times 10^{-4}$ \\     \cline{3-6}
                      & & $\phantom{0}2$  & $1.00 \times 10^{-3}$ &  $5.59 \times 10^{-4}$ &  $3.87 \times 10^{-4}$ \\     \cline{3-6}
                      & & $\phantom{0}4$  & $1.00 \times 10^{-3}$ &  $5.68 \times 10^{-4}$ &  $3.99 \times 10^{-4}$ \\     \cline{3-6}
                      & & $17$            & $1.00 \times 10^{-3}$ &  $5.40 \times 10^{-4}$ &  $3.74 \times 10^{-4}$ \\     \cline{3-6}
                      & & $19$            & $1.20 \times 10^{-3}$ &  $7.19 \times 10^{-4}$ &  $4.37 \times 10^{-4}$ \\     \cline{3-6}
                      & & $23$            & $1.20 \times 10^{-3}$ &  $7.12 \times 10^{-4}$ &  $4.64 \times 10^{-4}$ \\     \cline{3-6}
                      & & $27$            & $1.30 \times 10^{-3}$ &  $7.20 \times 10^{-3}$ &  $5.17 \times 10^{-4}$ \\     \cline{3-6}
                      & & $34$            & $1.30 \times 10^{-3}$ &  $7.45 \times 10^{-3}$ &  $5.44 \times 10^{-4}$ \\     \cline{3-6}
                      & & $42$            & $1.50 \times 10^{-3}$ &  $8.00 \times 10^{-3}$ &  $6.04 \times 10^{-4}$ \\     \cline{2-6}
                & LMMSE & $17$            & $2.46 \times 10^{-1}$ &  $1.01 \times 10^{-1}$ &  $1.65 \times 10^{-1}$\\  \hline
\end{tabular}
\end{center}
\end{table}

Table \ref{tb:1bit_num_MSE_compare} also shows that the proposed
algorithm can only achieve \emph{reasonable} performance with
$\mathcal{K}= 17$. This finding naturally raises the question: How
many quantization bits of these $17$ measurements are needed to
achieve a performance \emph{close to} that of the unquantized
measurements? Therefore, Table \ref{tb:1bit_model_test_result} shows
the performance of the proposed algorithm using different
quantization bits for the $17$ measurements, where the number of
bits used for quantization is denoted as $\mathcal{B}$. For ease of
reference, the performance of the proposed algorithm with the
unquantized measurements is also provided. Furthermore, the average
running time of the proposed algorithm for Table
\ref{tb:1bit_model_test_result} is provided in Table
\ref{tb:run_time_diff_bit_69bus}. Table
\ref{tb:1bit_model_test_result} shows that increasing $\mathcal{B}$
results in the improvement of the state estimation performance.
However, as shown in Table \ref{tb:run_time_diff_bit_69bus}, the
required running time also increases with the value of
$\mathcal{B}$. Fortunately, the required running time is within 2 s
even at $\mathcal{B}=6$. Moreover, if we further increase the
quantization bit from $\mathcal{B} = 6$ to $\mathcal{B} = 7$, the
performance remains the same. However, the corresponding running
time rapidly increases from 1.90 s to 2.45 s. These findings
indicate that $(\mathcal{B},\mathcal{K}) = (6,17)$ are appropriate
parameters for the proposed framework. Therefore, in the following
simulations, we consider the scenario where more than half of the
measurements are quantized to $6$ bits and the others are
unquantized to reduce the data transmitted from the measuring
devices to the data gathering center further.

\begin{table}
\begin{center}
\caption{The performance of {\tt EMSwGAMP} using different quantization bits for the $17$ measurements for the various systems}\label{tb:1bit_model_test_result}
\begin{tabular}{|c|c|c|c|c|c|}
   \hline
   \multirow{2}{*}{$N$} & \multirow{2}{*}{$\mathcal{B}$-bit} & \multirow{2}{*}{${\rm MSE}$} & \multirow{2}{*}{${\rm MSE}_{\rm magn}$} & \multirow{2}{*}{${\rm MSE}_{\rm phase}$} \\
                        &                           && &  \\  \hline\hline
   \multirow{7}{*}{69\phantom{m}} & 1-bit    & $9.49 \times 10^{-4}$ &  $5.40 \times 10^{-4}$ &  $4.05 \times 10^{-4}$ \\    \cline{2-5}
                       & 2-bit    & $9.49 \times 10^{-4}$ &  $5.40 \times 10^{-4}$ &  $4.05 \times 10^{-4}$ \\    \cline{2-5}
                       & 3-bit    & $9.47 \times 10^{-4}$ &  $5.39 \times 10^{-4}$ &  $4.04 \times 10^{-4}$ \\    \cline{2-5}
                       & 4-bit    & $9.11 \times 10^{-4}$ &  $5.22 \times 10^{-4}$ &  $3.87 \times 10^{-4}$ \\    \cline{2-5}
                       & 5-bit    & $8.80 \times 10^{-4}$ &  $5.06 \times 10^{-4}$ &  $3.71 \times 10^{-4}$ \\    \cline{2-5}
                       & 6-bit    & $8.69 \times 10^{-4}$ &  $5.00 \times 10^{-4}$ &  $3.66 \times 10^{-4}$ \\    \cline{2-5}
                       & unquantized       & $3.78 \times 10^{-4}$ &  $2.02 \times 10^{-4}$ &  $1.75 \times 10^{-4}$ \\    \hline \hline
 \multirow{7}{*}{69m}  & 1-bit    & $9.79 \times 10^{-4}$ &  $5.53 \times 10^{-4}$ &  $3.91 \times 10^{-4}$ \\    \cline{2-5}
                       & 2-bit    & $9.79 \times 10^{-4}$ &  $5.53 \times 10^{-4}$ &  $3.91 \times 10^{-4}$ \\    \cline{2-5}
                       & 3-bit    & $9.75 \times 10^{-4}$ &  $5.49 \times 10^{-4}$ &  $3.89 \times 10^{-4}$ \\    \cline{2-5}
                       & 4-bit    & $9.36 \times 10^{-4}$ &  $5.28 \times 10^{-4}$ &  $3.73 \times 10^{-4}$ \\    \cline{2-5}
                       & 5-bit    & $9.04 \times 10^{-4}$ &  $5.12 \times 10^{-4}$ &  $3.58 \times 10^{-4}$ \\    \cline{2-5}
                       & 6-bit    & $8.93 \times 10^{-4}$ &  $5.06 \times 10^{-4}$ &  $3.54 \times 10^{-4}$ \\    \cline{2-5}
                       & unquantized       & $7.15 \times 10^{-4}$ &  $3.09 \times 10^{-4}$ &  $3.74\times 10^{-4}$ \\    \hline
\end{tabular}
\end{center}
\end{table}

\begin{table}
\begin{center}
\caption{The run time of {\tt EMSwGAMP} using different quantization bits for the $17$ measurements for the various systems}\label{tb:run_time_diff_bit_69bus}
\begin{tabular}{|c|c|c|c|c|c|c|}
   \cline{1-3} \cline{5-7}
   \multirow{2}{*}{$N$} & \multirow{2}{*}{$\mathcal{B}$-bit} & \multirow{2}{*}{Time (s)} & & \multirow{2}{*}{$N$} & \multirow{2}{*}{$\mathcal{B}$-bit} & \multirow{2}{*}{Time (s)} \\
    & & & & & &　\\ \cline{1-3} \cline{5-7}   \cline{1-3} \cline{5-7}
   \multirow{7}{*}{69\phantom{m}}                   & 1-bit    & $1.22$           & &\multirow{7}{*}{69m} & 1-bit    & $1.17$ \\    \cline{2-3} \cline{6-7}
                                                    & 2-bit    & $1.28$           & &                     & 2-bit    & $1.23$ \\    \cline{2-3} \cline{6-7}
                                                    & 3-bit    & $1.31$           & &                     & 3-bit    & $1.29$ \\    \cline{2-3} \cline{6-7}
                                                    & 4-bit    & $1.37$           & &                     & 4-bit    & $1.36$ \\    \cline{2-3} \cline{6-7}
                                                    & 5-bit    & $1.56$           & &                     & 5-bit    & $1.53$ \\    \cline{2-3} \cline{6-7}
                                                    & 6-bit    & $1.90$           & &                     & 6-bit    & $1.87$ \\    \cline{2-3} \cline{6-7}
                                                    & unquantized       & $0.20$  &  &                    & unquantized       & $0.20$\\
  \cline{1-3} \cline{5-7}
\end{tabular}
\end{center}
\end{table}

Table \ref{tb:34_42_q_six_bits} shows the average performance of two
algorithms with $\mathcal{K}=34$ and $\mathcal{K}=42$ for the two
test systems. Here, $\mathcal{K}$ denotes the number of 6-bit
quantized measurements and the performance of {\tt EMSwGAMP} using
only the unquantized measurements is also listed for convenient
reference. Notably, the proposed {\tt EMSwGAMP} algorithm
significantly outperforms LMMSE, where the performance of LMMSE
deteriorates again to an unacceptable level. We also observed that
increasing the number of $6$ bit quantized measurements from
$\mathcal{K}=34$ to $\mathcal{K}=42$ only results in a slight
performance degradation for {\tt EMSwGAMP}. Consequently, by using
{\tt EMSwGAMP}, we can drastically reduce the amount of transmitted
data \emph{without} compromising performance. The total amount of
measurements of a 69-bus test system is $76$, where $68$ current
measurements and 8 voltage measurements originate from the meters
and PMUs, respectively. Therefore, if the measurements are quantized
as $16$ bits for the conventional meters and PMUs, $16 \times 76 =
1, 216$ bits should be transmitted. However, for the proposed
algorithm with $\mathcal{K} = 34$ (i.e., 34 measurements quantized
with 6 bits and 42 measurements quantized with 16 bits), only $34
\times 6 + 42 \times 16 = 876$ bits should be transmitted. In this
case, the transmission data can be reduced by $27.96\%$. Similarly,
for the proposed algorithm with $\mathcal{K} = 42$ (i.e., 42
measurements quantized with 6 bits and 34 measurements quantized
with 16 bits), the transmission data can be reduced by $34.53\%$. In
addition, we further discuss the required transmission bandwidth of
the proposed framework and the conventional system under the
assumption that the meters can update measurements every 1 s. As
defined by IEEE 802.11n, when the data are modulated with quadrature
phase-shift keying for a 20 MHz channel bandwidth, the data rate is
$21.7$ Mbps. Therefore, we can approximate the transmission rate as
1.085/Hz/s. For the proposed algorithm with $\mathcal{K} = 34$ and
$\mathcal{B} = 6$ (i.e.,  $876$ bits should be transmitted), the
required transmission bandwidth is $808$ Hz. However, the required
transmission bandwidth for the conventional system (i.e., $1,216$
bits should be transmitted) is $1, 121$ Hz. In this case, the
proposed architecture can also reduce the transmission bandwidth by
$27.83\%$. Notably, this study only focuses on the reduction of the
transmission data. However, references that specifically discuss the
smart meter data transmission system are few (e.g.,
\cite{DN-12-CM,2016-smartgrid-comm-meter}), which are not considered
in this study. Further studies can expand the scope of  the present
work  to include these transmission mechanism to provide more
efficient transmission framework.

\begin{table}
\begin{center}
\caption{Average estimation results obtained by {\tt EMSwGAMP} and LMMSE with different numbers of $6$-bit quantized measurements for the various systems}\label{tb:34_42_q_six_bits}
\begin{tabular}{|C{0.35cm}|c|C{1.25cm}|C{1.45cm}|C{1.45cm}|C{1.45cm}|}
   \hline
   \multirow{2}{*}{$N$} &\multirow{2}{*}{$\mathcal{K}$} & \multirow{2}{*}{Algorithm} & \multirow{2}{*}{${\rm MSE}$} & \multirow{2}{*}{${\rm MSE}_{\rm magn}$} & \multirow{2}{*}{${\rm MSE}_{\rm phase}$} \\
    & &&& &  \\  \hline\hline
   \multirow{6}{*}{69\phantom{m}} & \multirow{3}{*}{34} &  unquantized              & $3.86 \times 10^{-4}$ &  $2.11 \times 10^{-4}$ &  $1.78 \times 10^{-4}$      \\     \cline{3-6}
                         &                       & {\tt EMSwGAMP} & $1.00 \times 10^{-3}$ &  $6.01 \times 10^{-4}$ &  $4.34 \times 10^{-4}$    \\     \cline{3-6}
                         &                       & LMMSE          & $9.39 \times 10^{-1}$ &  $9.34 \times 10^{-1}$ &  $2.21 \times 10^{-1}$    \\     \cline{2-6}
                         & \multirow{3}{*}{42}   &  unquantized              & $3.77 \times 10^{-4}$ &  $2.06 \times 10^{-4}$ & $1.72 \times 10^{-4}$   \\     \cline{3-6}
                         &                       & {\tt EMSwGAMP} & $7.89 \times 10^{-4}$ &  $4.52 \times 10^{-4}$ &  $3.32 \times 10^{-4}$    \\     \cline{3-6}
                         &                       & LMMSE          & $9.39 \times 10^{-1}$ &  $9.32 \times 10^{-1}$ &  $2.88 \times 10^{-1}$    \\     \hline
   \multirow{6}{*}{69m} & \multirow{3}{*}{34} &  unquantized             & $7.78 \times 10^{-4}$ &  $3.89 \times 10^{-4}$ & $3.56 \times 10^{-4}$    \\    \cline{3-6}
                         &                       & {\tt EMSwGAMP} & $1.10 \times 10^{-3}$ &  $6.28 \times 10^{-4}$ &  $4.31 \times 10^{-4}$    \\     \cline{3-6}
                         &                       & LMMSE          & $1.02$ &  $1.02$ &  $2.10 \times 10^{-1}$    \\        \cline{2-6}
                         & \multirow{3}{*}{42} &  unquantized            & $6.77 \times 10^{-4}$ &  $2.86 \times 10^{-4}$ & $3.59 \times 10^{-4}$    \\    \cline{3-6}
                         &                       & {\tt EMSwGAMP} & $1.20 \times 10^{-3}$ &  $6.56 \times 10^{-4}$ &  $4.53 \times 10^{-4}$    \\     \cline{3-6}
                         &                       & LMMSE          & $1.02$ &  $1.02$ &  $2.69 \times 10^{-1}$    \\    \hline
\end{tabular}
\end{center}
\vspace{-.15in}
\end{table}

Finally, we evaluate the performance of the proposed {\tt EMSwGAMP}
algorithm for three-phase state estimation. The above simulation
results show that almost half of the measurements can be represented
with low-precision. Hence, in this test system, $\mathcal{K}=51$
measurements are quantized with 6-bit. Therefore, Table
\ref{tb:48_q_six_bits} shows the average performance of two
algorithms with $\mathcal{K}=51$ for IEEE 37-bus three-phase system.
The performance of {\tt EMSwGAMP} using only unquantized
measurements is also included for ease of reference. Table
\ref{tb:48_q_six_bits} shows that {\tt EMSwGAMP} still outperformed
LMMSE but only with a slight degradation as compared to the
unquantized result. The proposed {\tt EMSwGAMP} algorithm can reduce
transmission data by $25.91\%$ compared to the high-precision
measurement data. Hence, the proposed algorithm can be applied not
only to a single-phase but also to a three-phase system. Most
importantly, the proposed algorithm can also decrease the amount of
data required to be transmitted and processed.

\begin{table}
\begin{center}
\caption{Average estimation results obtained through {\tt EMSwGAMP}
and LMMSE with $6$-bit quantized measurements for the three-phase
systems}\label{tb:48_q_six_bits}
\begin{tabular}{|C{0.35cm}|c|C{1.25cm}|C{1.45cm}|C{1.45cm}|C{1.45cm}|}
   \hline
   \multirow{2}{*}{$N$} &\multirow{2}{*}{$\mathcal{K}$} & \multirow{2}{*}{Algorithm} & \multirow{2}{*}{${\rm MSE}$} & \multirow{2}{*}{${\rm MSE}_{\rm magn}$} & \multirow{2}{*}{${\rm MSE}_{\rm phase}$} \\
    & &&& &  \\  \hline\hline
   \multirow{3}{*}{37} & \multirow{3}{*}{51} &  unquantized             & $2.02 \times 10^{-4}$ &  $1.17 \times 10^{-4}$ & $8.79 \times 10^{-5}$    \\    \cline{3-6}
                         &                       & {\tt EMSwGAMP} & $6.64 \times 10^{-4}$ &  $4.42 \times 10^{-4}$ &  $2.30 \times 10^{-4}$    \\     \cline{3-6}
                         &                       & LMMSE          & $0.92$ &  $0.92$ &  $1.49 \times 10^{-4}$    \\        \cline{1-6}

\end{tabular}
\end{center}
\end{table}

\section{Conclusion}\label{sec:05}
We first proposed a data reduction technique via coarse quantization
of partial uncensored measurements and then developed a new
framework based on a Bayesian belief inference to incorporate
quantization-caused measurements of different qualities to obtain an
optimal state estimation and reduce the amount of data while still
incorporating different quality of data. The simulation results
indicated that the proposed algorithm performs significantly better
than other linear estimates, even for a case scenario in which more
than half of measurements are quantized to $6$ bits. This finding
verifies the effectiveness of the proposed scheme.

\appendices
\section{How the measurements are being
picked for quantization} For ease of explanation, a 69-bus test
system is provided in Fig. 4, where the subset of buses
$\mathcal{M}=\{1,2,\ldots,27\}$, called the \emph{main} chain of the
system, plays an important role for estimating the system states.
Therefore, the measurements from the \emph{side} chain of the system
are selected for quantization. In addition, the numbers of the
quantized measurements considered in this study are $\mathcal{K} =
2$, $\mathcal{K}=4$, $\mathcal{K} = 17$, $\mathcal{K} = 19$,
$\mathcal{K} = 23$, $\mathcal{K} = 27$, $\mathcal{K}=34$, and
$\mathcal{K}=42$. More specifically, if $\mathcal{K} = 2$, the
current measurements from the subset of buses $\{12,68,69\}$ are
selected for quantization; if $\mathcal{K} = 4$, the current
measurements from the subset of buses $\{12,68,69\}$ and
$\{11,66,67\}$ are selected for quantization; if $\mathcal{K} = 17$,
the current measurements from the subset of buses $\{12,68,69\}$,
$\{11,66,67\}$ and $\{9,53,54,\ldots,65\}$ are being picked for
quantization; if $\mathcal{K} = 19$, the current measurements from
the subset of buses $\{12,68,69\}$, $\{11,66,67\}$, $\{8,51,52\}$
and $\{9,53,54,\ldots,65\}$ are selected for quantization; if
$\mathcal{K} = 23$, the current measurements from the subset of
buses $\{12,68,69\}$, $\{11,66,67\}$, $\{8,51,52\}$,
$\{4,47,48,49,50\}$, and $\{9,53,54,\ldots,65\}$ are selected for
quantization; if $\mathcal{K} = 27$, the current measurements from
the subset of buses $\{12,68,69\}$, $\{11,66,67\}$, $\{8,51,52\}$,
$\{3,28,29,\ldots,35\}$, and $\{9,53,54,\ldots,65\}$ are selected
for quantization; if $\mathcal{K} = 34$, the current measurements
from the subset of buses $\{12,68,69\}$, $\{11,66,67\}$,
$\{8,51,52\}$, $\{4,47,48,49,50\}$, $\{36,38,\ldots,46\}$, and
$\{9,53,54,\ldots,65\}$ are selected for quantization; if
$\mathcal{K} = 42$, the current measurements from the subset of
buses $\{12,68,69\}$, $\{11,66,67\}$, $\{8,51,52\}$,
$\{4,47,48,49,50\}$, $\{3,28,29,\ldots,35\}$, $\{36,38,\ldots,46\}$,
and $\{9,53,54,\ldots,65\}$ are selected for quantization.

\begin{figure}
\begin{center}
\resizebox{3.5in}{!}{%
\includegraphics*{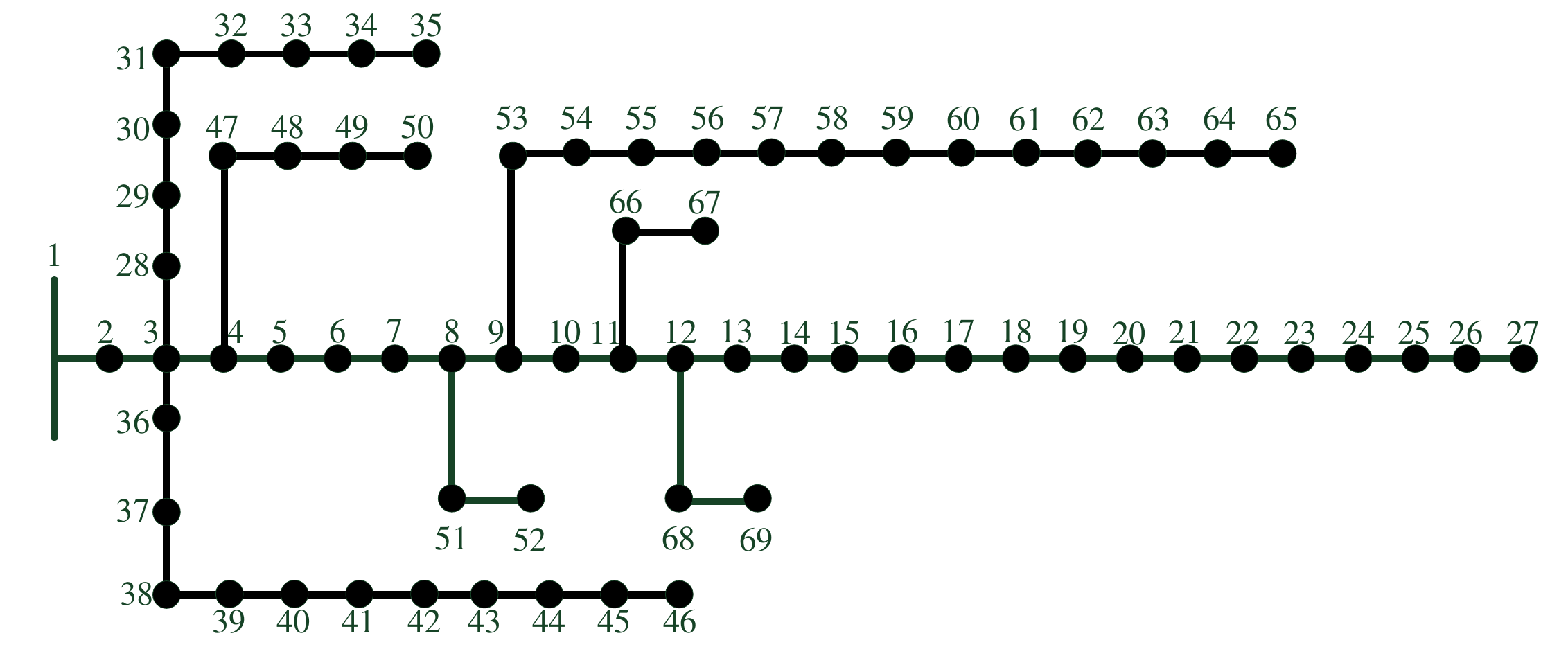} }%
\caption{A 69-bus test system.}\label{fig:69bus}
\end{center}
\end{figure}

\bibliographystyle{IEEEtran}

\begin{thebibliography}{10}
\providecommand{\url}[1]{#1}
\csname url@samestyle\endcsname
\providecommand{\newblock}{\relax}
\providecommand{\bibinfo}[2]{#2}
\providecommand{\BIBentrySTDinterwordspacing}{\spaceskip=0pt\relax}
\providecommand{\BIBentryALTinterwordstretchfactor}{4}
\providecommand{\BIBentryALTinterwordspacing}{\spaceskip=\fontdimen2\font plus
\BIBentryALTinterwordstretchfactor\fontdimen3\font minus
  \fontdimen4\font\relax}
\providecommand{\BIBforeignlanguage}[2]{{%
\expandafter\ifx\csname l@#1\endcsname\relax
\typeout{** WARNING: IEEEtran.bst: No hyphenation pattern has been}%
\typeout{** loaded for the language `#1'. Using the pattern for}%
\typeout{** the default language instead.}%
\else
\language=\csname l@#1\endcsname
\fi
#2}}
\providecommand{\BIBdecl}{\relax}
\BIBdecl

\bibitem{SG-techreport}
``The smart grid: An introduction,'' prepared by Litos Strategic Communication
  for U.S. Department of Energy, Tech. Rep., Oct. 2008.

\bibitem{Huang-12}
Y.-F. Huang, S.~Werner, J.~Huang, N.~Kashyap, and V.~Gupta, ``State estimation
  in electric power grids: meeting new challenges presented by the requirements
  of the future grid,'' \emph{IEEE Signal Process. Mag.}, vol.~29, no.~5, pp.
  33--43, Sep. 2012.

\bibitem{Nagasawa-12}
K.~Nagasawa, C.~R. Upshaw, J.~D. Rhodes, C.~L. Holcomb, D.~A. Walling, and
  M.~E. Webber, ``Data management for a large-scale smart grid demonstration
  project in {Austin, Texas},'' in \emph{Proc. ASME Int. Conf. Energy
  Sustainability}, San Diego, CA, Jul. 2012, pp. 1027--1031.

\bibitem{abur2004powe}
A.~Abur and A.~G. Exposito, \emph{Power System State Estimation: Theory and
  Implementation}.\hskip 1em plus 0.5em minus 0.4em\relax Boca Raton, FL: CRC
  Press, 2004.

\bibitem{Li-14-TPS}
{X. Li, A. Scaglione, and T. H. Chang}, ``A framework for phasor measurement
  placement in hybrid state estimation via gauss」newton,'' \emph{IEEE Trans.
  Power Syst.}, vol.~29, no.~2, pp. 824--832, Mar. 2014.

\bibitem{Gol-14}
M.~G\"{o}l and A.~Abur, ``{LAV} based robust state estimation for systems
  measured by {PMUs},'' \emph{IEEE Trans. Smart Grid}, vol.~5, no.~4, pp.
  1808--1814, Jul. 2014.

\bibitem{Hurtgen-08}
M.~Hurtgen and J.-C. Maun, ``Advantages of power system state estimation using
  phasor measurement units,'' in \emph{Proc. Power Syst. Comput. Conf.},
  Glasgow, Scotland, Jul. 2008, pp. 1--7.

\bibitem{Phadke-09}
A.~G. Phadke, J.~S. Thorp, R.~F. Nuqui, and M.~Zhou, ``Recent developments in
  state estimation with phasor measurements,'' in \emph{Proc. IEEE Power Syst.
  Conf. Expo.}, Seattle, WA, Mar. 2009, pp. 1--7.

\bibitem{PMU-09-book}
A.~G. Phadke and J.~S. Thorp, \emph{{Synchronized Phasor Measurements and Their
  Aplication}}.\hskip 1em plus 0.5em minus 0.4em\relax New York: Springer,
  2008.

\bibitem{Zhou-06}
M.~Zhou, V.~A. Centeno, J.~S. Thorp, and A.~G. Phadke, ``An alternative for
  including phasor measurements in state estimators,'' \emph{IEEE Trans. Power
  Syst.}, vol.~21, no.~4, pp. 1930--1937, Nov. 2006.

\bibitem{Gol-15}
M.~G\"{o}l and A.~Abur, ``A hybrid state estimator for systems with limited
  number of {PMUs},'' \emph{IEEE Trans. Power Syst.}, vol.~30, no.~3, pp.
  1511--1517, May 2015.

\bibitem{Alam-14}
S.~M.~S. Alam, B.~Natarajan, and A.~Pahwa, ``Distribution grid state estimation
  from compressed measurements,'' \emph{IEEE Trans. Smart Grid}, vol.~5, no.~4,
  pp. 1631--1642, Jul. 2014.

\bibitem{Rangan-11-ISIT}
{S.~Rangan}, ``Generalized approximate message passing for estimation with
  random linear mixing,'' in \emph{Proc. IEEE Int. Symp. Inf. Theory}, Saint
  Petersburg, Russia, Aug. 2011, pp. 2168--2172.

\bibitem{Krzakala-12}
F.~Krzakala, M.~M\'{e}zard, F.~Sausset, Y.~Sun, and L.~Zdeborov\'{a},
  ``Probabilistic reconstruction in compressed sensing: Algorithms, phase
  diagrams, and threshold achieving matrices,'' \emph{J. Stat. Mech. Theory
  Exp.}, vol. 2012, no.~8, Aug. 2012, {A}rt. ID P08009.

\bibitem{SwAMP-2015}
{A.~Manoel, F.~Krzakala, E.~W.~Tramel, and L.~Zdeborov\'{a}}, ``Swept
  approximate message passing for sparse estimation,'' in \emph{Proc. 32nd Int.
  Conf. Mach. Learn.}, Lille, France, Jul. 2015, pp. 1123--1132.

\bibitem{Vila-13TSP}
{J.~P.~Vila and P.~Schniter}, ``{Expectation-maximization Gaussian-mixture
  Approximate Message Passing},'' \emph{IEEE Trans. Sig. Proc.}, vol.~61,
  no.~19, pp. 4658--4672, Oct. 2013.

\bibitem{Chen-91-TPD}
T.-H. Chen, M.-S. Chen, K.-J. Hwang, P.~Kotas, and E.~Chebli, ``Distribution
  system power flow analysis---{A} rigid approach,'' \emph{IEEE Trans. Power
  Del.}, vol.~6, no.~3, pp. 1146--1152, Jul. 1991.

\bibitem{Jones-11}
K.~D. Jones, ``Three-phase linear state estimation with phasor measurements,''
  Master's thesis, Elect. Comput. Eng. Dept., Virginia Polytech. Inst. State
  Univ., Blacksburg, VA, USA, May 2011.

\bibitem{Proakis-book}
J.~G. Proakis and M.~Salehi, \emph{Digital Communications}, 5th~ed.\hskip 1em
  plus 0.5em minus 0.4em\relax New York, USA: McGraw-Hill, 2008.

\bibitem{Wen-15}
{C.-K. Wen~{\it et al.}}, ``Bayes-optimal joint channel-and-data estimation for
  massive {MIMO} with low-precision {ADCs},'' \emph{IEEE Trans. Signal
  Process.}, vol.~64, no.~10, pp. 2541--2556, May 2016.

\bibitem{Hu-11-CIM}
{Y. ~Hu, A. ~Kuh, A. ~Kavcic, and T. ~Yang}, ``{A Belief Propagation Based
  Power Distribution System State Estimator},'' \emph{IEEE Comput. Intell.
  Mag.}, vol.~6, no.~3, pp. 36--46, Aug. 2011.

\bibitem{Pea-book-88}
J.~Pearl, \emph{Probabilistic Reasoning in Intelligent Systems}, 2nd~ed.\hskip
  1em plus 0.5em minus 0.4em\relax San Francisco, CA: Kaufmann, 1988.

\bibitem{FG-01-IT}
{F.~R.~Kschischang, B.~J.~Frey, and H.-A.~Loeliger}, ``Factor graphs and the
  sum-product algorithm,'' \emph{IEEE Trans. Inf. Theory}, vol.~47, no.~2, pp.
  498--519, Feb. 2001.

\bibitem{MP-book}
C.~M. Bishop, \emph{Pattern Recognition and Machine Learning}.\hskip 1em plus
  0.5em minus 0.4em\relax New York, NY, USA: Springer, 2006.

\bibitem{Guo-06-ITW}
{D.~Guo and C.~C.~Wang}, ``A symptotic mean-square optimality of belief
  propagation for sparse linear systems,'' in \emph{Proc. IEEE Inf. Theory
  Workshop}, Chengdu, China, Oct. 2006, pp. 194--198.

\bibitem{Wang-14-WCNC}
{S.~Wang, Y.~Li, and J.~Wang}, ``Low-complexity multiuser detection for uplink
  large-scale {MIMO},'' in \emph{Proc. IEEE Wireless Commun. Netw. Conf.},
  Istanbul, Turkey, Apr. 2014, pp. 236--241.

\bibitem{Jay-15-TVT}
{J.-C.~Chen, C.-J.~Wang, K.-K.~Wong, and C.-K.~Wen}, ``Low-complexity precoding
  design for massive multiuser {MIMO} systems using approximate message
  passing,'' \emph{IEEE Trans. Veh. Technol.}, vol.~65, no.~7, pp. 5707--5714,
  Jul. 2016.

\bibitem{Wang-15-ICC}
{S.~Wang, Y.~Li, and J.~Wang}, ``Large-scale antenna system with massive
  one-bit iintegrated energy and information receivers,'' in \emph{Proc. IEEE
  Int. Conf. Commun.}, London, UK, Jun. 2015, pp. 2024--2029.

\bibitem{Barbier-15}
J.~Barbier, C.~Sch\"{u}lke, and F.~Krzakala, ``Approximate message-passing with
  spatially coupled structured operators, with applications to compressed
  sensing and sparse superposition codes,'' \emph{J. Stat. Mech. Theory Exp.},
  vol. 2015, no.~5, May 2015, {A}rt. ID P05013.

\bibitem{69-bus-data}
{R.~Paras}, ``{Load Flow Analysis of Radial Distribution Network using Linear
  Data Structure},'' \emph{arXiv preprint arXiv:1403.4702}, 2014.

\bibitem{MATPOWER}
{R.~D.~Zimmerman, C.~E.~Murillo-S\'{a}nchez, and R.~J.~Thomas}, ``{MATPOWER}
  steady-state operations, planning and analysis tools for power systems
  research and education,'' \emph{IEEE Trans. Power Syst.}, vol.~26, no.~1, pp.
  12--19, Feb. 2011.

\bibitem{DN-12-CM}
{D. Niyato and P. Wang}, ``Cooperative transmission for meter data collection
  in smart grid,'' \emph{IEEE Commun. Mag.}, vol.~50, no.~4, pp. 90--97, Apr.
  2012.

\bibitem{2016-smartgrid-comm-meter}
{C.~Karupongsiri, K.~S.~Munasinghe, and A.~Jamalipour}, ``A novel communication
  mechanism for smart meter packet transmission on {LTE} networks,'' in
  \emph{Proc. IEEE Int. Conf. Smart Grid Comm.}, Sydney, Australia, Nov. 2016,
  pp. 1--6.

\end{thebibliography}

\end{document}